\documentclass[11pt,a4paper]{article}

\pdfoutput=1
\usepackage{jheppub}
\usepackage{simplewick}
\usepackage{thumbpdf}
\usepackage{amsmath}
\usepackage{amssymb}
\usepackage{dsfont}
\usepackage{array}
\usepackage{cancel}
\usepackage[hang,small]{caption2}
\usepackage{subfigure}
\usepackage{slashed} 
\usepackage{verbatim}

\newcommand{\be}{\begin{equation}}
\newcommand{\ee}{\end{equation}}

\def\L {{L_0}}

\title{Surprises from the resummation of ladders in the ABJ(M) cusp anomalous dimension}

\author{Marisa Bonini$^{\bf a}$,}
\author{Luca Griguolo$^{\bf a}$,}  
\author{Michelangelo Preti$^{\bf a}$}
\author{and Domenico Seminara$^{\bf b}$}
\affiliation{$^{\bf a}$ Dipartimento di Fisica e Scienze della Terra, Universit\`a di Parma and INFN Gruppo Collegato di Parma, Viale G.P. Usberti 7/A, 43100 Parma, Italy}
\affiliation{$^{\bf b}$ Dipartimento di Fisica, Universit\`a di Firenze and INFN Sezione di Firenze, via G. Sansone 1, 50019 Sesto Fiorentino, Italy}
\emailAdd{marisa.bonini@fis.unipr.it} 
\emailAdd{luca.griguolo@fis.unipr.it} 
\emailAdd{michelangelo.preti@fis.unipr.it} 
\emailAdd{domenico.seminara@fi.infn.it} 

\abstract{We study the cusp anomalous dimension in ${\cal N}=6$ ABJ(M) theory, 
identifying a scaling limit in which the ladder diagrams dominate. The resummation is 
encoded into a Bethe-Salpeter equation that is mapped to a Schroedinger problem, exactly 
solvable due to the surprising supersymmetry of the effective Hamiltonian. In the ABJ case the 
solution implies the diagonalization of the $U(N)$ and $U(M)$ building blocks, suggesting the 
existence of two independent cusp anomalous dimensions and an unexpected exponentiation 
structure for the related Wilson loops. While consistent with previous perturbative analysis, the 
strong coupling limit of our result does not agree with the string theory computation, 
emphasizing a difference with the analogous resummation in the ${\cal N}=4$ case.}

\begin{document}
\maketitle

\section{Introduction and summary of the results}
The duality between ${\cal N}=6$ Super Chern-Simons theory with matter (also known as ABJ(M)
 theory \cite{Aharony:2008ug,Aharony:2008gk}) and string theory on $AdS_4\times CP_3$ represents one the most interesting 
 possibilities to explore AdS/CFT correspondence beyond the original paradigm. 
 Although it seems to share many similarities with the cousin ${\cal N}=4$ Super Yang-Mills 
 theory in four dimensions, there are still many aspects calling for a better comprehension. 
 Supersymmetric Wilson loops, in particular, provide a rich class of BPS observables \cite{Gaiotto:2007qi}-\cite{Berenstein:2008dc} 
 that can be computed exactly through localization technique in the simplest situations \cite{Kapustin:2009kz}. 
 While their quantum behavior is still rather mysterious in the general case \cite{Cardinali:2012ru}, the 
 well-understood 1/2 BPS and 1/6 BPS circles exhibit an intriguing non-trivial interpolation 
 between weak and strong coupling regime \cite{Marino:2009jd}-\cite{Marino:2011eh}. A careful study of the relevant matrix models \cite{Marino:2011eh} has also unveiled a  variety of phenomena of string/M-theory origin \cite{Hatsuda:2012dt}-\cite{Hatsuda:2013oxa}. 
  This contrasts with the relative simplicity of  ${\cal N}=4$ SYM, where a gaussian matrix 
  model \cite{Erickson:2000af,Drukker:2000rr} describes exactly the dynamics of 1/2 BPS Wilson loops \cite{Pestun:2007rz}. 

Wilson loops are also relevant in gauge theories (supersymmetric or not) because they encode important properties of scattering amplitudes and infrared radiation: the cusp anomalous dimension $\Gamma(\varphi)$, originally introduced in \cite{Polyakov:1980ca} as the ultraviolet divergence of a Wilson loop with Euclidean cusp angle $\varphi$, appears in fact in many interesting physical situations. It was computed in QCD to two-loop order in \cite{Korchemsky:1987wg} and, in light-like limit $\varphi\to ~i\infty$, provides crucial universal informations \cite{Korchemsky:1992xv}. In supersymmetric theories $\Gamma(\varphi)$ is not a BPS observable, making its exact computation in ${\cal N}=4$ SYM a difficult challenge. In the light-like limit
\cite{Beisert:2006ez} integrability controls its all-order behavior through an integral equation, whose solution matches both weak coupling expansions \cite{Bern:2006ew} and string computations at the strong coupling \cite{Gubser:2002tv}.  In the general case a strategy \cite{Correa:2012hh,Drukker:2012de} for computing  $\Gamma(\varphi)$ was later proposed (see \cite{Bombardelli:2009ns,Gromov:2009bc,Arutyunov:2009ur} for the original introduction of the related TBA approach in the case of local operators). The cusp anomalous dimension can be generalized including an $R$-symmetry angle $\theta$ that distinguishes the coupling of the scalars to the 
 two halves of the cusp \cite{Drukker:2011za}. The new observable $\Gamma(\varphi,\theta)$ interpolates between BPS configurations and generalized quark-antiquark potentials. Exact equations can be written applying integrability and they have been checked successfully at three loops \cite{Correa:2012nk,Correa:2012hh}. In the near-BPS limit it is possible to use localization to obtain the exact form of the so-called Bremsstrahlung function \cite{Correa:2012at,Fiol:2012sg}, that has been later directly recovered from the TBA equations \cite{Gromov:2012eu,Gromov:2013qga}. 
More recently localization has been used to derive the Bremsstrahlung function in presence of local operator insertions \cite{Bonini:2015fng}
and in more general superconformal field theories \cite{Fiol:2015spa}. 

As anticipated before, even for the simple circular 1/2 BPS Wilson loop the weak/strong 
interpolation for ABJ(M) theory is not non-trivial. Integrability itself has been explored here 
in a somehow limited range of situations \cite{oai:arXiv.org:0806.3951}-\cite{oai:arXiv.org:0807.0777}: when established it still depends on an 
elusive interpolating function, $h(\lambda)$ \cite{oai:arXiv.org:0806.3391}-\cite{oai:arXiv.org:0806.4959}. Recently a proposal for the functional form of 
$h(\lambda$) has been advanced \cite{Gromov:2014eha} and checked at two-loop level in string theory \cite{Bianchi:2014ada}, under suitable assumptions. Alternatively $h(\lambda)$ could be determined by computing exactly some quantity by integrability and confronting 
with the same calculation by localization (or by other QFT techniques in which unknown 
functions are absent). A natural candidate would be the ABJM Bremsstrahlung function, for which 
all order proposals exist \cite{Bianchi:2014laa}, but integrability has not been yet applied to its determination.

For this reason we think it is important to study $\Gamma(\varphi,\theta)$ 
in ${\cal N}=6$ Super Chern-Simons theory: it would be useful to obtain exact QFT results to 
be compared with the integrability approach and, at strong coupling, with string theory. 
$\Gamma(\varphi,\theta)$ has been introduced in ABJ(M) theory in \cite{Griguolo:2012iq}, where its 
computation at two-loop has been presented and its exponentiation properties discussed. 
The two halves of the cusp are locally 1/2 BPS and therefore couple also directly to the 
fermions of the theory, as originally discovered in \cite{Drukker:2009hy}, and not only to the gauge connections 
and scalars. The resulting cusped Wilson loop is not globally supersymmetric and its 
exact evaluation seems very challenging. Fortunately, the analogous system in ${\cal N}=4$ 
SYM can be tackled in a particular limit through Feynman diagrams resummation. In \cite{Correa:2012nk} 
a new scaling limit involving the complexified angle $\theta$ was introduced,
\begin{equation}\label{scalN4}
i\theta \gg 1, \,\,\,\,\,\,\, \lambda \ll 1, \,\,\,\,\,\, \hat{\lambda} =\lambda \exp(i\theta/4)\,\,\,{\rm fixed.} 
\end{equation}
Here $\lambda = g^2N$ is the 't Hooft coupling of ${\cal N}=4$ SYM. In this limit  the leading order contribution is simply given by ladder diagrams, where the rungs are made by scalar exchanges. 
The ladder diagrams can be summed up efficiently using a Bethe-Salpeter equation, solved 
exactly in the small $\varphi$ limit. The strong coupling behavior has been also examined, 
finding agreement with the corresponding string theory calculation \cite{Correa:2012nk}. Later it was also 
performed an analysis at next-to-leading order, generalizing the original Bethe-Salpeter 
equation and computing the relevant corrections at strong coupling \cite{Henn:2012qz}. Remarkably 
these corrections have been also obtained from string theory and successfully compared 
each other. As repeatedly stressed in the original analysis \cite{Correa:2012nk}, the matching of the strong coupling limit of the Bethe-Salpeter solution with the string theory computation is quite surprising. The 
ladders limit, $\lambda\to 0$ with $\hat{\lambda}$ fixed, is different from the strong coupling 
limit $\lambda\to\infty$ with $i\theta\gg 1$ fixed and the result could, in principle, depend 
on their order: nevertheless they agree at leading and subleading level.

In this paper we consider a similar limit in three-dimensional ABJ(M) theory, 
\begin{equation}
\label{lim1}
i\theta \gg 1, \,\,\,\,\,\,\, \lambda_r \ll 1, \,\,\,\,\,\, \hat{\lambda}_r =\lambda_r\cos\frac{\theta}{2}, ~~~~ (r=1,2)
\end{equation}
obtaining some exact results for $\Gamma(\varphi, \hat{\lambda}_i)$. 
Here $\lambda_1 = \frac{N}{k},\lambda_2= \frac{M}{k}$ are the 't Hooft couplings of the ABJ(M) theory with gauge group 
$U(N)\otimes U(M)$, while $k$ is the Chern-Simons level. The presence of 
fermionic couplings to the cusped loop inherits a surprising 
supersymmetric structure in the relevant Bethe-Salpeter equation. More precisely, the 
effective Schroedinger problem, associated to the  integral equation that resums planar diagrams 
in ${\cal N}=6$ Super Chern-Simons, enjoys an unexpected quantum mechanical supersymmetry.
 Because only the ground state matters in determining $\Gamma(\varphi, \hat{\lambda}_i)$ \cite{Correa:2012nk}, 
 supersymmetry produces an exact expression for $any$ value of the opening angle $\varphi$. This 
 is in sharp contrast with the ${\cal N}=4$ case, where an analytic solution for the Bethe-Salpeter
  equation exists only at $\varphi=0$ (that in this case is the only supersymmetric point of the 
  associated Schroedinger equation \cite{Correa:2012nk}).

In the ABJM case ($N=M$) we get a very simple solution: $\Gamma(\varphi, \hat{\lambda})$ 
is exact at one-loop level, as in an abelian theory. The delicate balance between bosonic and 
fermionic contributions, encoded into the effective supersymmetric quantum mechanics, 
exponentiates without non-abelian correction the one-loop term. As a matter of fact we 
do not observe any transition between a weak-coupling and a strong-coupling regime and we cannot match our result with semiclassical computations in string 
theory, suggesting that it should exist a problem with the order of limits in this case.

In the ABJ case ($N\neq M$) the story is even more intriguing: the Bethe-Salpeter equation 
reduces to two coupled integral equations, resumming contributions from the upper ($N\times N$) and the 
lower ($M\times M$) blocks of the holonomy of the $U(N|M)$ superconnection defining the Wilson loop \cite{Drukker:2009hy}. By diagonalizing the system we 
end up with a two-dimensional supersymmetric Schroedinger equation. From the knowledge of its ground state energy and wave-function we reconstruct the original cusped Wilson loops. This works in full generality in three dimensions with explicit UV an IR cut-offs on the 
lines, while using dimensional regularization we get an explicit solution only for 
$\varphi=0$. The main result of our investigation is that, in the ABJ case, the original cusped Wilson 
loop, the trace of the holonomy of the superconnection defined in \cite{Drukker:2009hy}, does not renormalize multiplicatively but it mixes, at quantum level,  with the supertrace. As a consequence we end up with two independent cusp anomalous dimensions, 
related to the renormalization constants of the operator eigenstates (with respect to the mixing). 
On the other hand we could have expected this fact from the very beginning: taking the trace 
of the holonomy of the superconnection is required to preserve the global supersymmetry of the 1/2 BPS Wilson 
loop \cite{Drukker:2009hy}, In our case supersymmetry is generically broken and we observe a mixing between the trace and (say) the supertrace. Again the cusp anomalous dimensions are exact at one-loop 
level 
\begin{equation}\begin{split}
\Gamma^{(1)}_{\text{cusp}}(\varphi)=&\frac{\sqrt{\hat{\lambda}_1\hat{\lambda}_2}}{\cos \frac{\varphi}{2}}\,,\\
\Gamma^{(2)}_{\text{cusp}}(\varphi)=&-\frac{\sqrt{\hat{\lambda}_1\hat{\lambda}_2}}{\cos 
\frac{\varphi}{2}}\,.
\end{split}\end{equation}
We expect that subleading corrections should change the simple pattern 
$\Gamma^{(2)}_{\text{cusp}}=-\Gamma^{(1)}_{\text{cusp}}$: we will come back on this point in the conclusions.

The paper is structured in the following way:  in Section \ref{S2} we introduce the cusped 
Wilson loop in ABJ(M) theory as the trace of a superconnection, following \cite{Griguolo:2012iq}. 
The limit in which the ladder diagrams dominate is described in Section \ref{S3}, where we also discuss 
how to derive the cusp anomalous dimensions both in the explicit cut-off scheme and in 
dimensional regularization. In Section \ref{sec:BSgeneral} we derive the relevant Bethe-Salpeter 
equation and, after diagonalization, we obtain the associated Schroedinger equation. We solve the supersymmetric Schroedinger equations for 
generic opening angle ${\varphi}$, in the cut-off scheme, and for $\varphi=0$ in dimensional 
regularization. In Section \ref{sec5} we obtain the cusp anomalous dimensions at leading order and discuss the operator mixing for the cusped Wilson loops. Our conclusions and the future 
directions to improve our results appear in Section \ref{conclu}. Two appendices complete our presentation.

\vskip 5pt
\section{The 1/2 BPS generalized cusped Wilson line in ABJ(M) theory}
 \label{S2}
In this Section we review the construction of supersymmetric Wilson lines in ABJ(M) theory \cite{Drukker:2009hy}. In $d=3$ the generalized gauge connection can be 
defined in two different ways according to the degree of preserved 
supersymmetry. Indeed, we can consider a purely bosonic gauge 
connection whose holonomy is,  for a suitable choice of the path,  1/6 BPS.
On the other hand, adding on the lines local couplings to the fermions, we can interpret 
the Wilson operator as the holonomy of a $U(N|M)$  superconnection obtaining, 
for the infinite straight line, a 1/2 BPS operator. Its peculiar structure was also investigated 
via the  so-called Higgsing procedure which gives a
physical explanation for the appearance of the superconnection \cite{Lee:2010hk}.

\subsection{The 1/2 BPS Wilson line}

In ABJ(M) theory the gauge sector consists of  two  gauge fields $(A_\mu)_{i}^{\ j}$ and
$(\hat{A}_\mu)_{\hat{i}}^{\ \hat{j}}$ belonging respectively  to the adjoint of $U(N)$ and $U(M)$.
 The matter sector contains the complex fields $(C_I)_{i}^{\ \hat{i}}$ and 
 $({\bar C}^{I})_{\hat{i}}^{\  i}$ as well as the fermions $(\psi_I)_{\hat{i} }^{ \ i}$ 
 and $({\bar\psi}^{I})_{i }^{\ \hat{i} }$. 
The fields $(C,\bar \psi)$ transform in  the $({\bf N},{\bf \bar M})$  of the gauge group
 $U(N)\times U(M)$ while the fields $(\bar C, \psi)$ live in the $({\bf \bar N},{\bf M})$. 
 The additional  capital index $I=1,2,3,4$  belongs to the $R$-symmetry group $SU(4)$.

The central idea of \cite{Drukker:2009hy} is to replace the $U(N)\times U(M)$ gauge connection with the 
super-connection
\begin{equation}
  \label{superconnection}
 \mathcal{L}(\tau) \equiv -i \begin{pmatrix}
i\mathcal{A}
&\sqrt{\frac{2\pi}{\kappa}}  |\dot x | \eta_{I}\bar\psi^{I}\\
\sqrt{\frac{2\pi}{\kappa}}   |\dot x | \psi_{I}\bar{\eta}^{I} &
i\hat{\mathcal{A}}
\end{pmatrix} \ \  \ \ \mathrm{with}\ \ \ \  \left\{\begin{matrix} \mathcal{A}\equiv A_{\mu} \dot x^{\mu}-\frac{2 \pi i}{\kappa} |\dot x| M_{J}^{\ \ I} C_{I}\bar C^{J}\\
\\
\hat{\mathcal{A}}\equiv\hat  A_{\mu} \dot x^{\mu}-\frac{2 \pi i}{\kappa} |\dot x| \hat M_{J}^{\ \ I} \bar C^{J} C_{I},
\end{matrix}\ \right.
\end{equation}
belonging to the super-algebra of $U(N|M)$.
In \eqref{superconnection} the 
coordinates $x^{\mu}(\tau)$ define the contour of the loop operator,  
while  $M_{J}^{\ \ I}$, $\hat M_{J}^{\ \ I}$  are the scalar couplings and 
$\eta_{I}^{\alpha}$,  $\bar{\eta}^{I}_{\alpha}$ are fermionic (2-components Grassmann 
even quantities)  ones.

The form of $\mathcal{L}$ is determined mainly by dimensional analysis and symmetry properties of the fields. 
In $d=3$ the scalars have classical dimension $1/2$, so they could only appear as bilinears, 
which are in the adjoint and therefore enter in the diagonal blocks together with the gauge fields. 
Instead the fermions have dimension 1 and should appear linearly. Since they transform in the 
bifundamental, they are naturally placed in the off-diagonal entries of the matrix.

For a given a path $\mathcal{C}$, it is possible to compute the holonomy of the superconnection 
\eqref{superconnection}
\begin{equation}\label{WLnontraced}
W[\mathcal{C}]\equiv \mathcal{P} \exp{\left(-i\int_{\mathcal{C}} 
\mathcal{L}(\tau)d\tau\right)}.
\end{equation}
When the contour is a straight-line $S$, all the couplings can be chosen to be independent of
$\tau$, i.e. constant, in order to preserve the invariance under translations along the 
line. Further restrictions on scalar and fermionic couplings follow from 
 R$-$symmetry  and supersymmetry requirements.  Actually, imposing 
 $\delta_{\text{susy}} \mathcal{L}(\tau)=0$ gives rise to  loop operators which are merely 
 bosonic ($\eta=\bar\eta=0$) and at most $1/6$ BPS \cite{Cardinali:2012ru}. The weaker condition 
of invariance under supersymmetry up to a super-gauge transformation brings out 
the $1/2$ BPS solution. For (finite) closed path one has to carefully consider the boundary 
conditions obeyed by the gauge functions to obtain a (super-)gauge invariant object.
For instance, in the circle case one has to take  the trace of \eqref{WLnontraced}.
For an infinite open circuit, such as the straight line, 
the naive statement that the fields vanish when $\tau=\pm \infty$  allows two 
possible supersymmetric operators
\begin{equation}\begin{split}\label{trstr}
\mathcal{W}_{-}=&\frac{1}{N-M}
\mathrm{Str}\left[\mbox{$\displaystyle\mathrm{P}\!\exp\left(-i \int d\tau 
\mathcal{L}(\tau)\right)$}\right]\;,\\
\mathcal{W}_{+}=&\frac{1}{N+M}
\mathrm{Tr}\left[\mbox{$\displaystyle\mathrm{P}\!\exp\left(-i \int d\tau 
\mathcal{L}(\tau)\right)$}\right].
\end{split}\end{equation}
Usually one mainly considers the second possibility, since, for particular angles, it is connected through a conformal transformation to a BPS closed loop \cite{Griguolo:2012iq}. Nevertheless the supertraced holonomy has a crucial role as well.

\subsection{The generalized cusp}\label{sec:S2}
\begin{figure}[!h]
\centering
	\includegraphics[width=8cm]{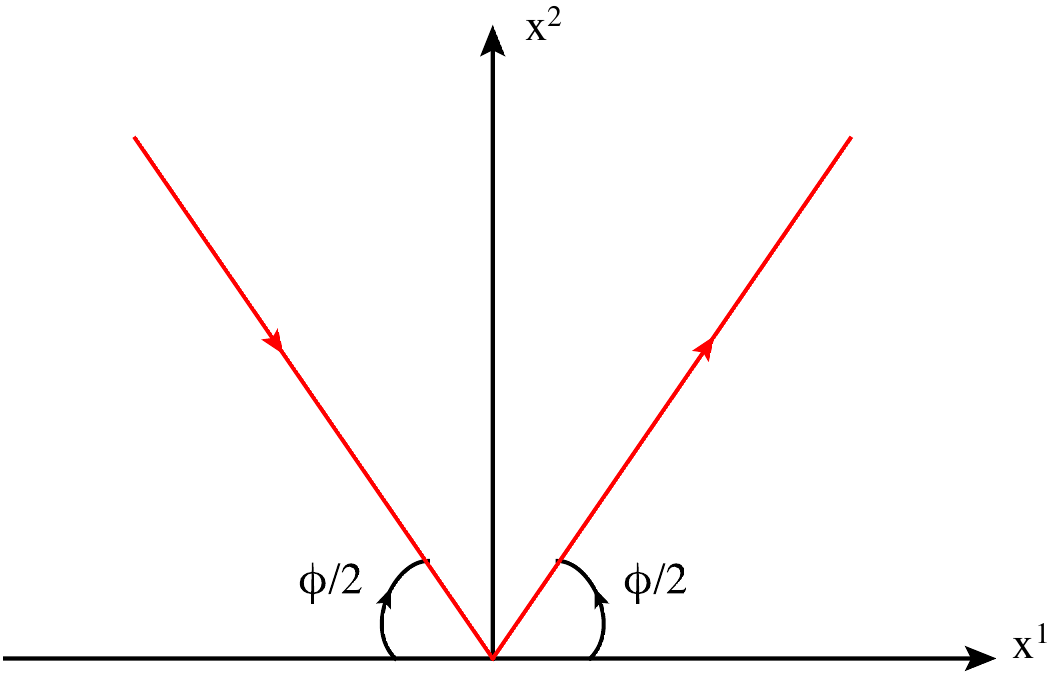}
	\caption{The planar cusp}
	\label{fig:cusp1}
\end{figure}
We consider the theory on the Euclidean space-time and take the contour depicted in 
Figure \ref{fig:cusp1}. 
The two rays  are in the plane $(1,2)$, intersect at the origin and are given by
\begin{equation}
x^\mu=\{0,\tau \cos \frac{\varphi}{2},|\tau| \sin \frac{\varphi}{2}\}\qquad 
-\infty\geq\tau\leq \infty\,.
\end{equation}
The angle between the rays is $\pi-\varphi$, thus $\varphi=0$ gives the continuous 
straight line. 

It is natural to consider  scalar and fermionic couplings 
different on the two segments of the cusp (but constant on each segment). 
The fermionic couplings have the factorized form 
\begin{equation}
\eta_{i M}^\alpha=n_{i M}\eta_i^\alpha\qquad \text{and}\qquad 
\bar{\eta}_{i\alpha}^M=n_{i}^M\bar{\eta}_{i\alpha}\,,
\end{equation}
where the index $i=1,2$ specifies which edge of the cusp we are 
considering. As discussed in \cite{Griguolo:2012iq} we can take:
\begin{equation}
\eta_{1\alpha} =\begin{pmatrix} 
e^{-i\frac{\varphi}{4}}& e^{i\frac{\varphi}{4}}
\end{pmatrix}\,, \qquad  
 \bar\eta_{1\alpha} =i
\begin{pmatrix} 
e^{i\frac{\varphi}{4}}\\ e^{-i\frac{\varphi}{4}}
\end{pmatrix}\,,
\end{equation}
and
\begin{equation}
\eta_{2\alpha} =\begin{pmatrix} 
e^{i\frac{\varphi}{4}}& e^{-i\frac{\varphi}{4}}
\end{pmatrix}\,, \qquad  
\bar\eta_{2\alpha} =i
\begin{pmatrix} 
e^{-i\frac{\varphi}{4}}\\ e^{i\frac{\varphi}{4}}
\end{pmatrix}\,.
\end{equation}
The $R-$symmetry part of the couplings is totally unconstrained and we choose 
\begin{equation}
n_{1M}=\mbox{\small$ \left(\cos\frac{\theta}{4}\ \ \sin\frac{\theta}{4}\ \ 0\ \ 0\right)$}
\qquad\text{and}\qquad
n_{2M}=\mbox{\small$ \left(\cos\frac{\theta}{4}\ \ -\sin\frac{\theta}{4}\ \ 0\ \ 
0\right)$}\,,
\end{equation}
(and we denote by  $\bar{n}_i^M$ the transpose of $n_{i M}$).
The matrices which couple the scalars on the two edges are 
\begin{equation}
M_{1J}^{\ \ I}=
\hat{M}_{1J}^{\ \ I}=\mbox{\small $\left(
\begin{array}{cccc}
 -\cos \frac{\theta }{2}& -\sin \frac{\theta }{2} & 0 & 0 \\
 -\sin \frac{\theta }{2}& \cos\frac{\theta }{2} & 0 & 0 \\
 0 & 0 & 1 & 0 \\
 0 & 0 & 0 & 1
\end{array}
\right)$}\ \ \ \ \mathrm{and}\ \ \ \  M_{2J}^{\ \ I}=
\hat{M}_{2J}^{\ \ I}=\mbox{\small $\left(
\begin{array}{cccc}
 -\cos \frac{\theta }{2} & \sin \frac{\theta }{2} & 0 & 0 \\
 \sin \frac{\theta }{2} & \cos\frac{\theta }{2} & 0 & 0 \\
 0 & 0 & 1 & 0 \\
 0 & 0 & 0 & 1
\end{array}
\right)$}.
\end{equation}
The quantum holonomy of the super-connection ${\cal L}$ 
in a representation ${\cal R}$ of the supergroup $U(N|M)$ is by definition
\begin{equation}
\label{loopexpectationvalue}
\left\langle \mathcal{W}_{\cal R} \right\rangle= \frac{1}{{\rm dim}_{\cal R}}\int {\cal D}[A,\hat{A},C,\bar{C},\psi,\bar{\psi}]~
{\rm e}^{-S_{\rm ABJ}}~{\rm Tr}_{\cal R} \left[
  {\rm P} \exp \left(- i\int_{\Gamma} d\tau\, {\cal L}(\tau) \right) \right],
\end{equation}
where $S_{ABJ}$ is the Euclidean action for $ABJ(M)$ theory (see Appendix \ref{sec:appA}). 
In the following $\mathcal{R}$ will be chosen to be the fundamental representation
of $U(N|M)$: this implies that the trace is taken in the fundamental 
representation $\textbf{N}$ or $\textbf{M}$ of the two gauge groups.

In general this Wilson loop operator is not supersymmetric unless $\theta=\pm \varphi$: in this case, having chosen the trace in its definition, it can be mapped by a suitable conformal transformation to a closed 1/6 BPS Wilson loop \cite{Griguolo:2012iq}.

\section{The cusp anomalous dimension in ABJ(M) theory and its computation through ladder diagrams}
\label{S3}

In this Section we discuss the definition of the cusp anomalous dimension in ABJ(M) 
theories and its computation in a limit in which ladder diagrams dominate. New features in the 
$N\neq M$ will emerge due to the exponentiation properties of the cusped loops.

\subsection{The cusp anomalous dimension in ABJ(M) theory}\label{sec:sezionemaledetta}

We start by recalling the $\mathcal{N}=4$ SYM case: the generalized cusp anomalous dimension is 
defined by the logarithmic divergent behaviour of a cusped Wilson loop \cite{Polyakov:1980ca}
\begin{equation}\label{definitiongamma}
\langle\mathcal{W}_{\text{cusp}}\rangle \simeq e^{-\Gamma_{\text{cusp}}(\varphi,\theta)
\log{\frac{\Lambda_{UV}}{\Lambda_{IR}}}}.
\end{equation}
Here $\Lambda_{UV}$ and $\Lambda_{IR}$ stand for the ultraviolet and infrared cut-offs respectively, that regularize the specific divergences associated to the cusp 
angle and the infinite extension of the lines. Typically one takes $\Lambda_{IR}=1/L$, where $L$ measures the (finite) length of the two edges 
and $\Lambda_{UV}=1/\delta$, with $\delta$ being a short-length scale either cutting or smoothing out 
the the cusp singularity. 
Alternatively one can use dimensional regularization \cite{Korchemsky:1987wg} and the logarithm is replaced by a simple pole $1/\epsilon$.
After the usual renormalization of the gauge theory\footnote{In ${\cal N}=4$ SYM this 
step is superfluous, being the $\beta$-function vanishing}, 
the relation between the bare and 
renormalized Wilson loop operator, for closed contours, is
\begin{equation}
\label{RenoW}
\mathcal{W}_{\text{cusp}}^{B}=Z_{\text{cusp}}\mathcal{W}^{R}_{\text{cusp}} \,,
\end{equation}
where the renormalization constant $Z_{\text{cusp}}$ depends on the dimensional regularization parameter $\epsilon$
and the subtraction point $\mu$ (in addition to the coupling constant 
$\lambda$).
When considering the standard cusp built out of two straight lines of length $L$, the relation \eqref{RenoW} must be, in general,
corrected as follows \cite{Aoyama:1981ev,Craigie:1980qs, Knauss:1984rx, Dorn:1986dt}
\be
\mathcal{W}_{\text{cusp}}^{B}=Z_{\text{cusp}} Z_{\text{open}}\mathcal{W}^{R}_{\text{cusp}}.
\ee
The second renormalization constant  cancels spurious divergences due to the fact that we are dealing with an open (and then non gauge-invariant) loop operator.  The separation between the two contributions is fixed by the renormalization condition $Z_{\text{cusp}}|_{\varphi=\theta=0}=1$.
The cusp anomalous dimension is defined as\footnote{In the definition of the anomalous dimension, the limit $\epsilon\rightarrow 0$ is understood.}
\begin{equation}\label{gammadef}
\Gamma_{\text{cusp}}(\varphi,\theta)=\mu \frac{d}{d\mu}\log{Z_{\text{cusp}}}\,,
\end{equation}
and plays the role of the anomalous 
dimension of a (non-local) quantum operator. 
Although the open cusped Wilson loop is not gauge invariant, the cusp anomalous dimension 
$\Gamma_{\text{cusp}}$ turns out to be so.
At the perturbative level the cusp divergence comes from
diagrams with propagators connecting both rays of the cusp and its exponentiation is governed 
by their maximal non-abelian part \cite{Korchemsky:1987wg}. 
In the $\mathcal{N}=4$ SYM case, the perfect balance between the gauge and the scalars contributions 
cancels, in the Feynman gauge, all the infinities related to integrations along 
the smooth part of the contour ($Z_{\text{open}}=1$).
Thus only the singularities associated to the discontinuous behavior at the cusp 
appear and one immediately singles out the relevant diagrams to be computed \cite{Makeenko:2006ds,Drukker:2011za}.

In the ABJ(M) case the situation is more subtle.
First of all the presence of the fermionic contributions breaks that balance 
and divergences persist even in the straight-line limit, at least in dimensional 
regularization and in the  Landau gauge \cite{Griguolo:2012iq}.
Moreover there are two gauge groups and the fermionic interactions, that live in the off-diagonal sector of the super-connection, mix non-trivially the $U(N)$ and $U(M)$ structures. 
Thus a non-standard form of exponentiation for the divergence of the cusped Wilson loop is expected.
Actually in the ABJM case $(N=M)$ the renormalization properties should be similar to the $\mathcal{N}=4$ SYM case, as explicitly checked at two-loop in \cite{Griguolo:2012iq}. In particular we will 
expect a single exponential behavior for the vacuum expectation value of the cusped loop operator.

In the ABJ case ($N\neq M$), as usual in the theory of renormalization of composed local operator,
we would expect instead the arising of a matrix-valued set of renormalization constants: 
\begin{equation}\label{mix1}
\mathcal{W}_a^B=\tilde{Z}_{ab}\,\mathcal{W}_b^R\,,
\end{equation}
where $a,b=\pm$ refers to the traced and supertraced operators\footnote{These two combinations provides a natural gauge invariant basis for the information contained in the superholonomy defined by $\mathcal{L}(\tau)$. The presence of these two different possibilities is the source of all the main differences with the $\mathcal{N}=4$ case, where only the trace makes sense.}.
Similarly to \eqref{gammadef}, the anomalous dimensions matrix is
\begin{equation}\label{gammageneral}
(\Gamma_{\text{cusp}})_{ab}=\left[\mu\frac{\partial}{\partial\mu}\log \tilde{Z}_{\text{cusp}}\right]_{ab}\,.
\end{equation}
In general the matrix  $\tilde{Z}_{\text{cusp}}$  in \eqref{gammageneral} is obtained  by rewriting the matrix  $\tilde{Z}_{ab}$ in \eqref{mix1} as $(Z_{\text{open}}\tilde{Z}_{\text{cusp}})_{ab}$ with normalization condition  $(\tilde{Z}_{\text{cusp}})_{ab}|_{\varphi=\theta=0}=\delta_{ab}$.
The scaling limit  \eqref{lim1} considered in this paper selects only diagrams 
connecting the two halves of the loop, then $Z_{\text{open}}=1$ and 
$\tilde{Z}=\tilde{Z}_{\text{cusp}}$.

\subsection{The scaling limit selecting ladder diagrams}
In \cite{Correa:2012nk} it was considered the scaling limit (\ref{scalN4}) in ${\cal N}=4$ SYM: there pure scalar exchanges between the rungs of the cusped Wilson loop become dominant and can be resummed by means of a Bethe-Salpeter equation \cite{Correa:2012nk}. Subleading corrections can be also systematically included in this scheme and consistency at strong coupling with semiclassical string computations has been found \cite{Henn:2012qz}.

In our case, we want to consider a similar limit: upon a quick inspection of the perturbative diagrams we recognize two types of relevant contributions (we refer to \cite{Griguolo:2012iq} for details on the perturbative expansion and related computations). At one-loop we have the single fermionic exchange (that is the same for the up and down diagonal blocks)

\begin{minipage}[r]{0.35\textwidth}
\centering{$\qquad\qquad\quad$\includegraphics[width=2.6cm]{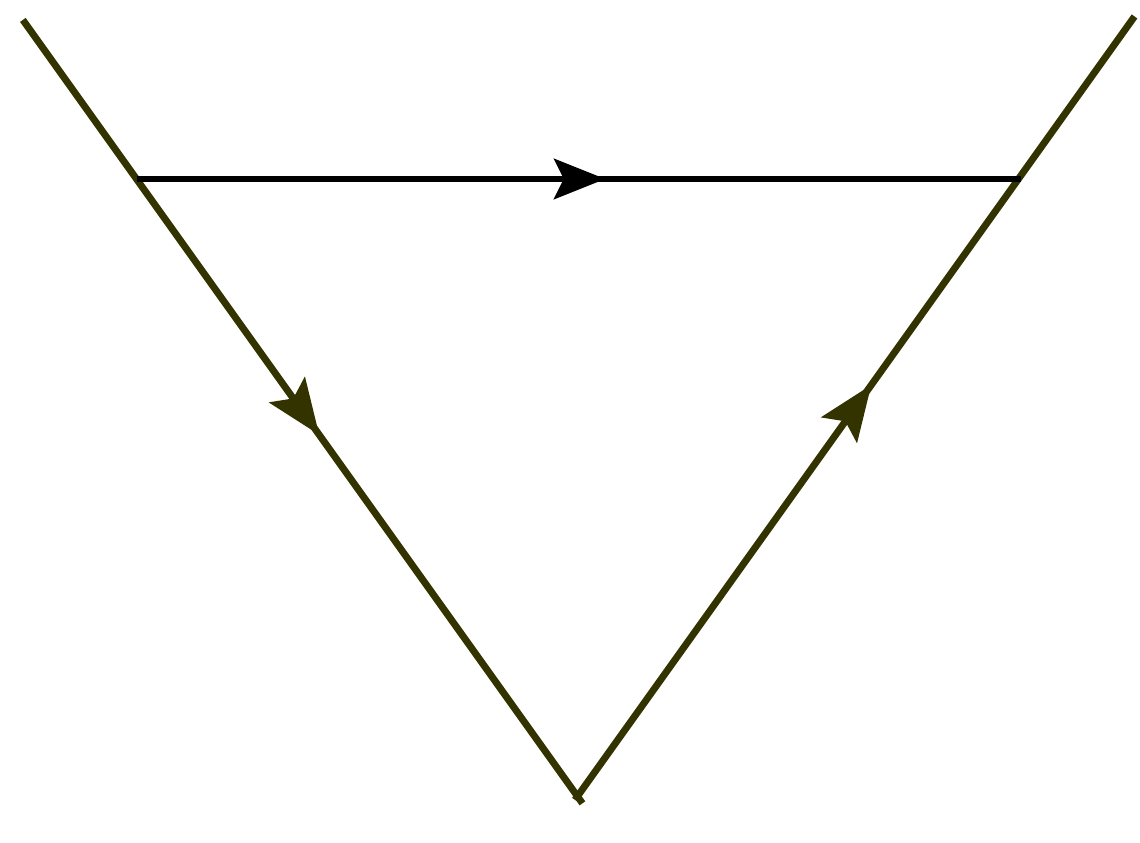}}
\end{minipage}%
\begin{minipage}[c]{0.2\textwidth}
\begin{center}\begin{equation*}
=\left(\frac{2\pi}{\kappa}\right)\frac{M N}{M+N}\frac{ \Gamma\left(\frac12-\epsilon\right)}{4\pi^{3/2-\epsilon}}
(\mu L)^{2\epsilon}\,\frac{1}{\epsilon}\frac{\cos\frac{\theta}{2}}{\cos\frac{\varphi}{2}}\,.
\end{equation*}\end{center}
\end{minipage}%

We have considered here the appropriate normalization of the trace. We notice that for $N=M$ this contribution is proportional to $\lambda\cos\frac{\theta}{2}$, suggesting to perform the 
scaling limit 
\begin{equation}\label{scalinglim1}
i\theta \gg 1, \,\,\,\,\,\,\, \lambda \ll 1, \,\,\,\,\,\, \hat{\lambda} =\lambda \cos\frac{\theta}{2}\,\,\,{\rm fixed.} 
\end{equation}
The natural generalization to the case $N\neq M$ is therefore
\begin{equation}\label{scalinglim}
i\theta \gg 1, \,\,\,\,\,\,\, \lambda_{1,2} \ll 1, \,\,\,\,\,\, \hat{\lambda}_{1,2} =\lambda_{1,2} \cos\frac{\theta}{2}\,\,\,{\rm fixed.} 
\end{equation}
At two-loop we observe that the above limit suppresses all the diagrams in which interactions are present. Obviously the double-fermionic exchange survives but also a pure scalar exchange comes into the game 

\begin{minipage}[r]{0.3\textwidth}
\centering{$\qquad\qquad$\includegraphics[width=3cm]{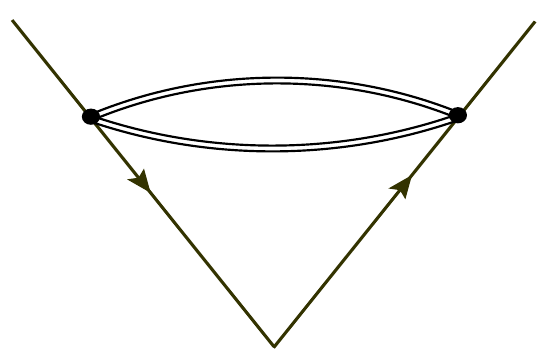}}
\end{minipage}%
\begin{minipage}[c]{0.20\textwidth}
\begin{equation*}
=-\left(\frac{2\pi}{\kappa}\right)^{\!\!2}MN\frac{\Gamma^{2}\!\left(\frac{1}{2}-\epsilon\right)}{16 \pi^{3-2\epsilon}}(\mu L)^{4\epsilon}
\cos^2\frac\theta 2\frac{1}{\epsilon}\frac{\varphi}{\sin\varphi}\,.
\end{equation*}
\end{minipage}%

This last contribution has exactly the same form of the one-loop scalar exchange in ${\cal N}=4$ SYM, except that here it appears at two-loop and the scaling behavior is different.

It is not difficult to realize that, at leading order, the generic diagrams surviving the limit 
consist of ladders made by fermionic and scalar exchanges, that should therefore summed 
up to obtain the complete result. We remark that the contributions coming from diagrams 
ending on a single line, and so leading to divergences not related to the cusp 
renormalization constant, are automatically suppressed in our limit. 
In the next Section we will derive an efficient way to sum up all the relevant ladder diagrams.

\section{Bethe-Salpeter equation for the generalized cusp in ABJ(M) at leading 
order in the scaling limit}\label{sec:BSgeneral}

\begin{figure}[!h]\begin{center}
\begin{minipage}[r]{0.25\textwidth}
\includegraphics[width=4cm]{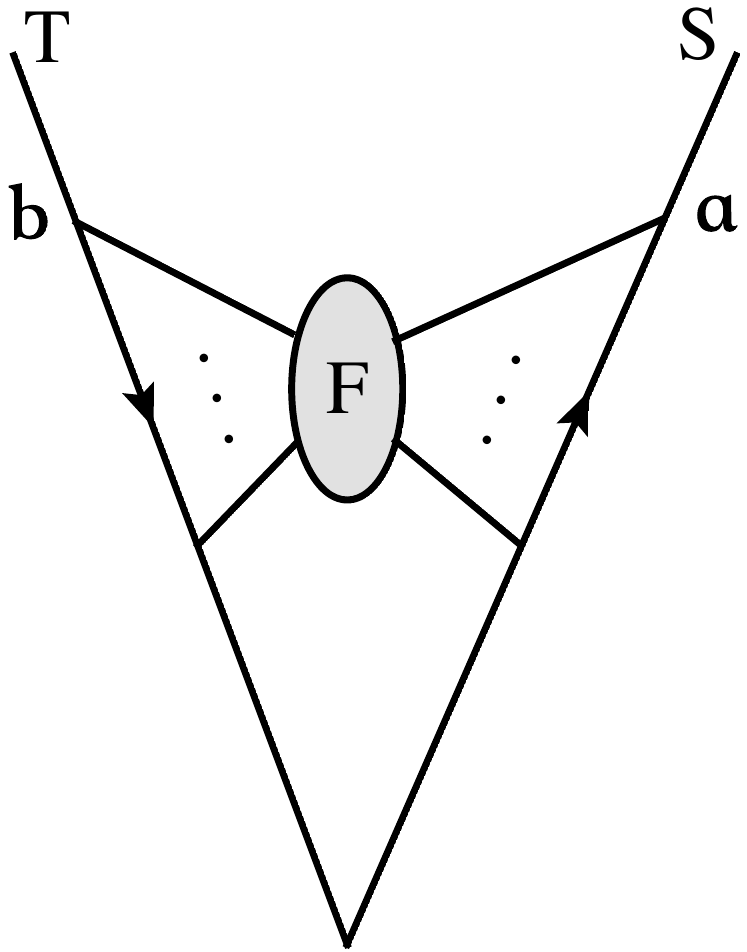}
\end{minipage}%
\begin{minipage}[c]{0.25\textwidth}
\Large\begin{equation*}
=\qquad\delta_a\,^b\qquad+
\end{equation*}
\normalsize
\end{minipage}%
\begin{minipage}[c]{0.25\textwidth}
\includegraphics[width=4cm]{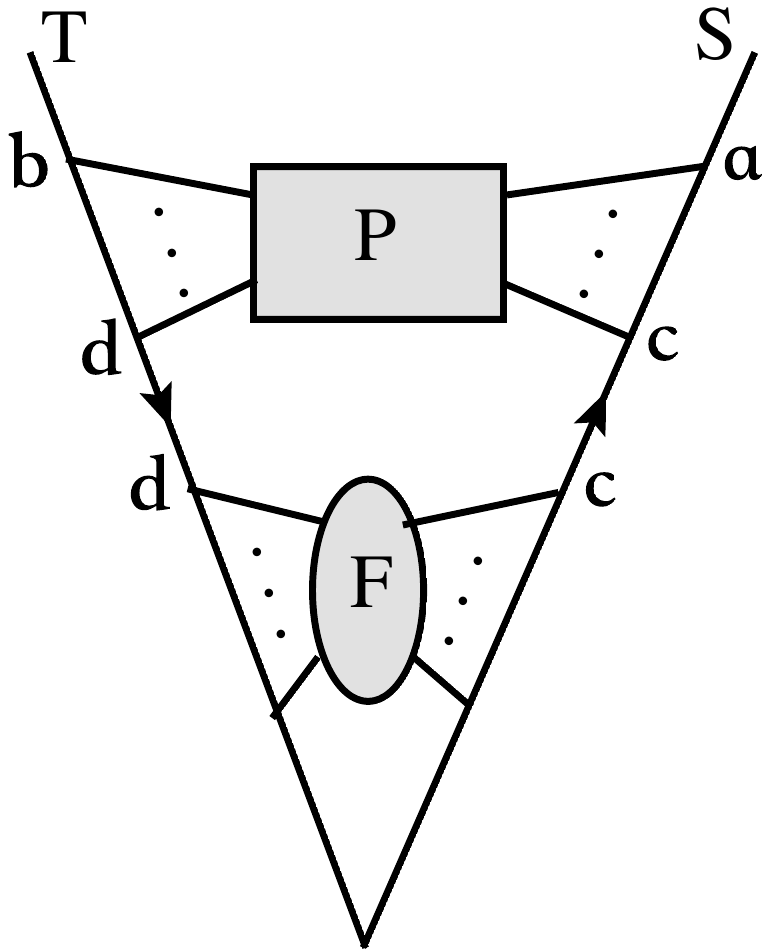}
\end{minipage}
\caption{Bethe-Salpeter equation at leading order}
\label{fig:BS}
\end{center}\end{figure}

The general strategy to sum up ladder diagrams for Wilson loops constructed by straight-lines was 
introduced in \cite{Erickson:1999qv} and used in \cite{Correa:2012nk} to derive the leading 
behavior of the ${\cal N}=4$ SYM cusp anomaly in the scaling limit. We adapt their technique 
to our case.

We denote the sum of the ladder diagrams with end-points within the intervals $(0,S)$ and $(0,T)$ of the cusp by $F_a\,^b(S,T)$, where $(a, b)$ are  group indices that can 
be $(i,j)\in U(N)$ or $(\hat{i},\hat{j})\in U(M)$,
respectively.
 $F_a\,^b(S,T)$ satisfies a Bethe-Salpeter equation
\begin{equation}\label{Bethe}
  F_a\,^b(S,T)=\delta_a\,^b+\int_0^S ds\int_0^T dt \;
  F_c\,^d(s,t) P_a\,^c\,_d\,^b(s,t)\,,
\end{equation}
that is shown schematically in Figure \ref{fig:BS}.
In the scaling limit \eqref{lim1} the scalar and the fermionic couplings of the loop 
dominate and one has to consider only exchanges of these fields between the two 
segments of the Wilson loop. The indices sequence follows the path-ordering of the Wilson loop and the scalar 
and fermionic propagators fix the kernel indices  as follow:
\begin{equation}
P_a\,^c\,_d\,^b(s,t)\simeq 
\delta_a\,^b\,\delta^c\,_d\times (\text{a ``scalar" function of $s$ and 
$t$}),
\end{equation}
in particular
\begin{equation}\begin{split}\label{indiciinizio}
&P_i\,^k\,_l\,^j(s,t)=M\delta_i\,^j\delta^k\,_lP^{(B)}(s,t)\,,\\
&P_i\,^{\hat{k}}\,_{\hat{l}}\,^j(s,t)=\delta_i\,^j\delta^{\hat{k}}\,_{\hat{l}}P^{(F)}(s,t)\,,\\
&P_{\hat{i}}\,^k\,_l\,^{\hat{j}}(s,t)=\delta_{\hat{i}}\,^{\hat{j}}\delta^k\,_lP^{(F)}(s,t)\,,\\
&P_{\hat{i}}\,^{\hat{k}}\,_{\hat{l}}\,^{\hat{j}}(s,t)=N\delta_{\hat{i}}\,^{\hat{j}}\delta^{\hat{k}}\,_{\hat{l}}P^{(B)}(s,t)\,,   
\end{split}\end{equation}
where $P^{(F)}(s,t)$ is the fermionic effective propagator and $P^{(B)}(s,t)$ the scalar 
effective propagator (double exchange) defined by \cite{Griguolo:2012iq}:
\begin{equation}\begin{split}
&P^{(F)}(s,t)=-\left(\frac{2\pi}{k}\right)\frac{\Gamma(1/2-\epsilon)\mu^{2\epsilon}}{4\pi^{3/2-\epsilon}}
\frac{\cos\theta/2}{\cos\varphi/2}\; 
(\partial_s+\partial_t)\frac{1}{(s^2+t^2+2st\cos\varphi)^{\frac{1}{2}-\epsilon}}\,,\\
&P^{(B)}(s,t)=\left(\frac{2\pi}{k}\right)^2\frac{\Gamma^2(1/2-\epsilon)\mu^{4\epsilon}}{4\pi^{3-2\epsilon}}
\frac{\cos^2\theta/2}{\cos^2\varphi/2}\;\frac{\cos^2\varphi/2}{(s^2+t^2+2st\cos\varphi)^{1-2\epsilon}}\,.
\end{split}\end{equation}
According to $U(N)$ or $U(M)$ indices \eqref{Bethe} splits into 
\begin{equation}
\begin{split}
&F_i\,^j(S,T)=\delta_i\,^j+\int_0^S ds\int_0^T dt \left(
 M F_k\,^l(s,t)\delta_i\,^j\delta^k\,_lP^{(B)}(s,t)+F_{\hat{k}}\,^{\hat{l}}(s,t)\delta_i\,^j\delta^{\hat{k}}\,_{\hat{l}}P^{(F)}(s,t),
 \right),\\
&F_{\hat{i}}\,^{\hat{j}}(S,T)=\delta_{\hat{i}}\,^{\hat{j}}+\int_0^S ds\int_0^T dt \left(
F_k\,^l(s,t)\delta_{\hat{i}}\,^{\hat{j}}\delta^k\,_lP^{(F)}(s,t)+N F_{\hat{k}}\,^{\hat{l}}(s,t)\delta_{\hat{i}}\,^{\hat{j}}\delta^{\hat{k}}\,_{\hat{l}}P^{(B)}(s,t)
\right).
\end{split}
\end{equation}
Thus, defining 
\begin{equation}\begin{split}
&F(S,T)=\frac{1}{\sqrt{N}}\text{Tr}_{\bf N}[
F_i\,^j(S,T)]\,,
\\
&\hat{F}(S,T)=\frac{1}{\sqrt{M}}\text{Tr}_{\bf 
M}[F_{\hat{i}}\,^{\hat{j}}(S,T)]\,,
\end{split}\end{equation}
we get
\begin{equation}\label{traceBS}
\begin{split}
&F(S,T)=\sqrt{N}+\int_0^S ds\int_0^T dt \left(
 MN F(s,t)P^{(B)}(s,t)+\sqrt{MN} \hat{F}(s,t)P^{(F)}(s,t)\right),\\
&\hat{F}(S,T)=\sqrt{M}+\int_0^S ds\int_0^T dt \left(
\sqrt{MN} F(s,t) P^{(F)}(s,t)+MN \hat{F}(s,t) P^{(B)}(s,t)\right).
\end{split}
\end{equation}

\begin{figure}[!h]\begin{center}
\begin{minipage}[r]{0.2\textwidth}
\includegraphics[width=3cm]{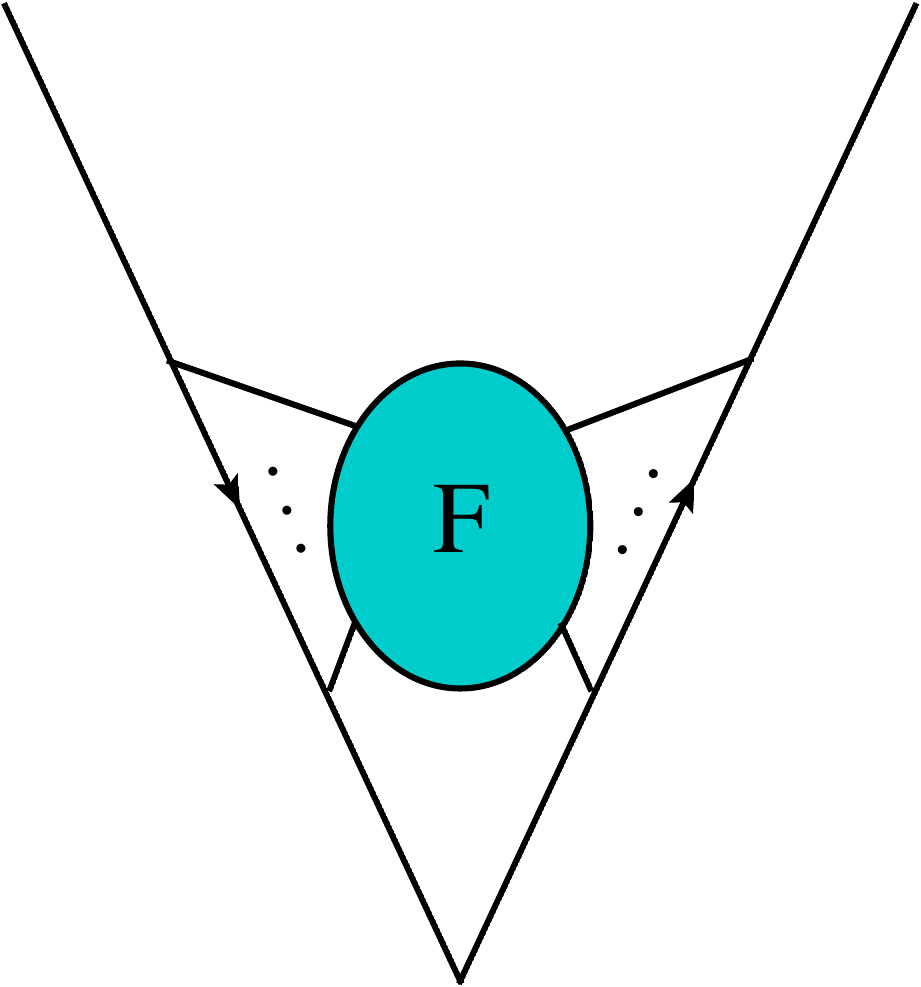}
\end{minipage}%
\begin{minipage}[c]{0.20\textwidth}
\Large\begin{equation*}
\!\!\!\!=~~\sqrt{N} ~+~MN\cdot~
\end{equation*}
\normalsize
\end{minipage}%
\begin{minipage}[c]{0.21\textwidth}
\includegraphics[width=3cm]{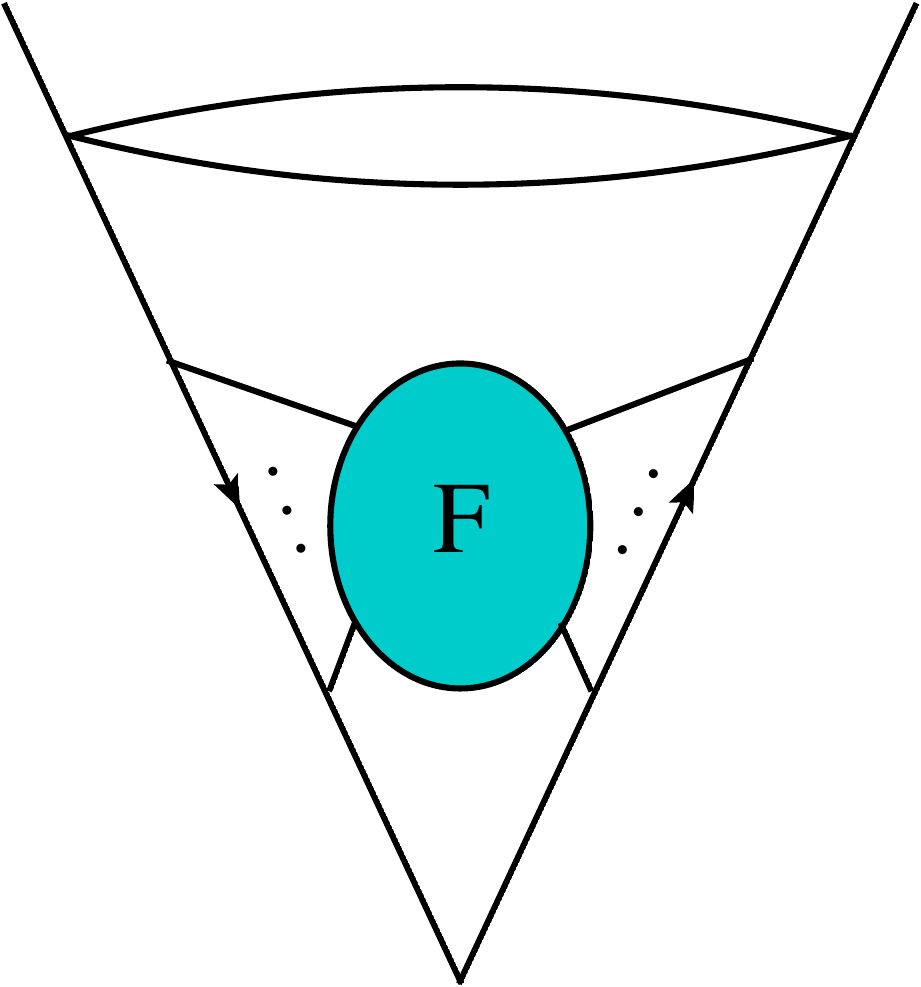}
\end{minipage}
\begin{minipage}[c]{0.09\textwidth}
\Large\begin{equation*}\!\!\!\!
+~~\sqrt{MN}~\cdot~
\end{equation*}
\normalsize
\end{minipage}%
\begin{minipage}[c]{0.225\textwidth}
\includegraphics[width=3cm]{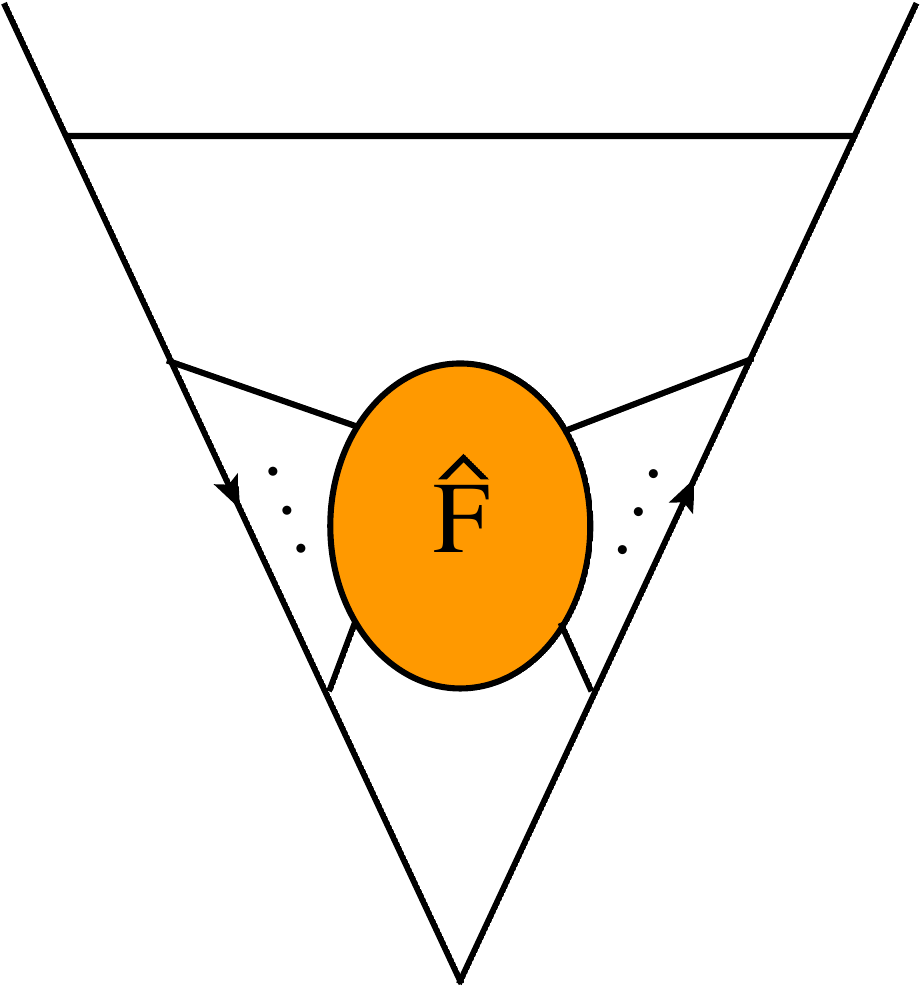}
\end{minipage}
\begin{minipage}[r]{0.2\textwidth}
\includegraphics[width=3cm]{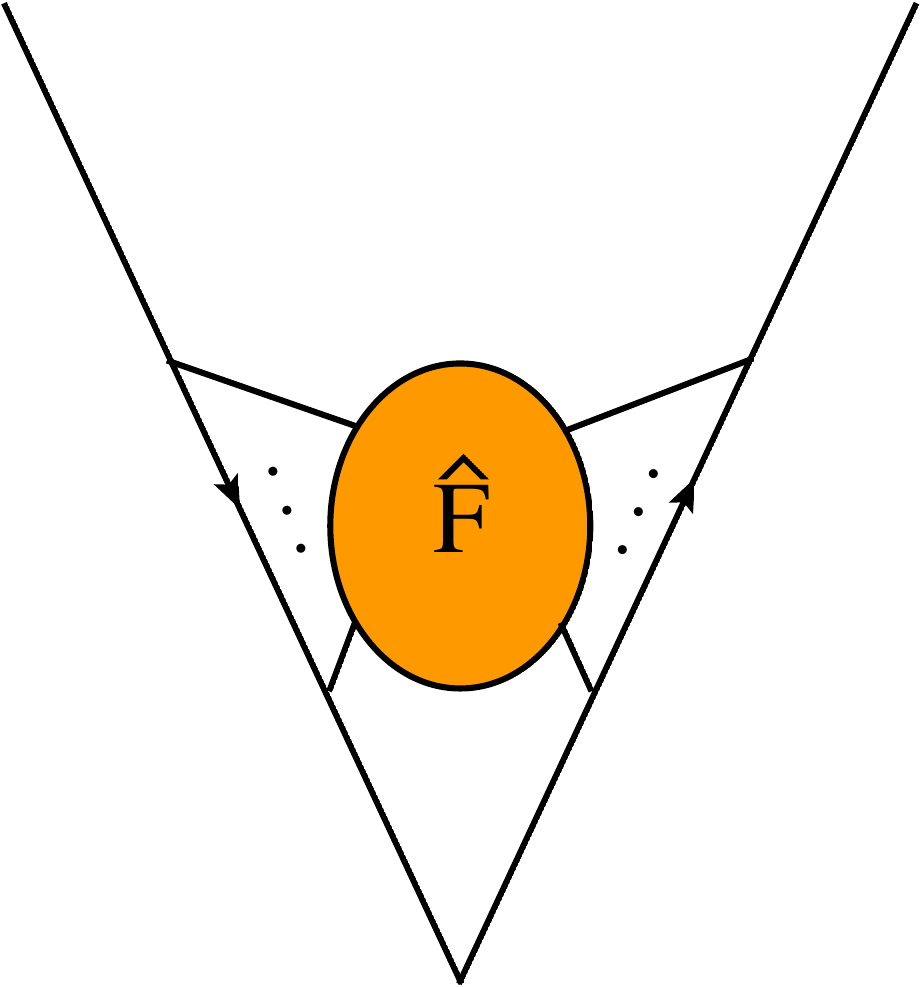}
\end{minipage}%
\begin{minipage}[c]{0.22\textwidth}
\Large\begin{equation*}
\!\!\!\!=~~\sqrt{M} +\sqrt{MN}~\cdot~
\end{equation*}
\normalsize
\end{minipage}%
\begin{minipage}[c]{0.20\textwidth}
\includegraphics[width=3cm]{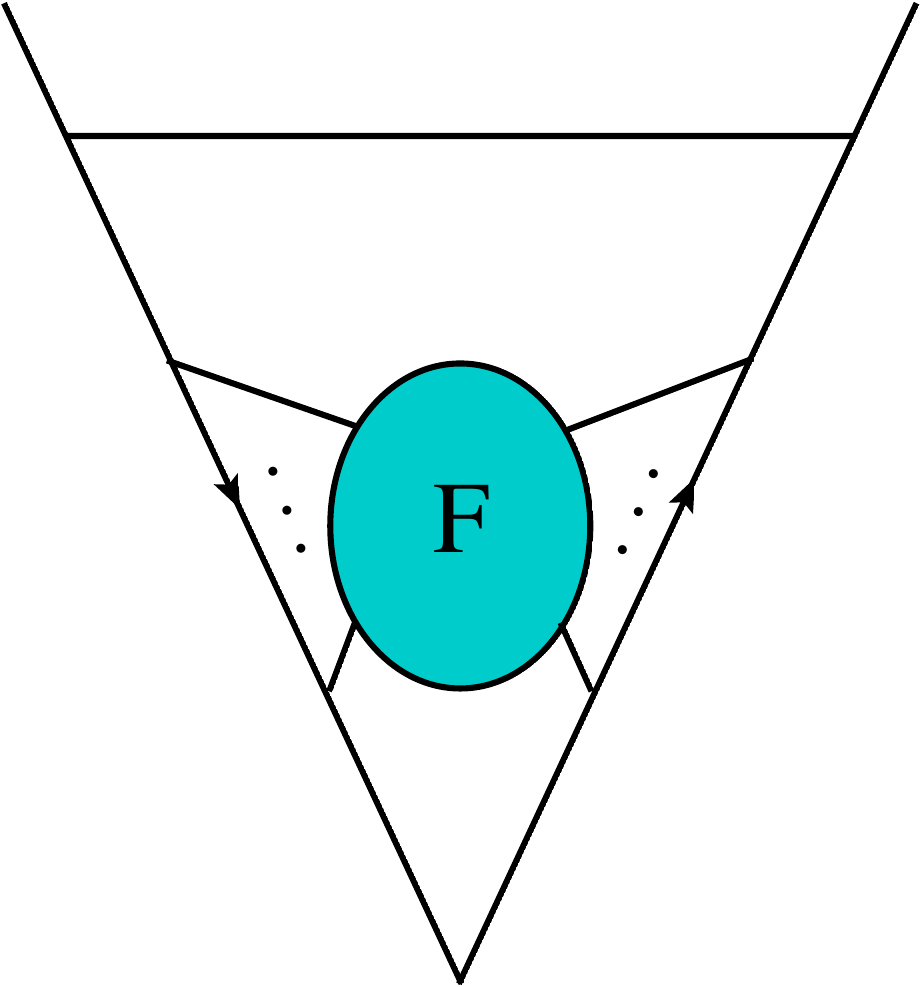}
\end{minipage}
\begin{minipage}[c]{0.07\textwidth}
\Large\begin{equation*}\!\!\!\!
+~~~ MN\cdot
\end{equation*}
\normalsize
\end{minipage}%
\begin{minipage}[c]{0.21\textwidth}
\includegraphics[width=3cm]{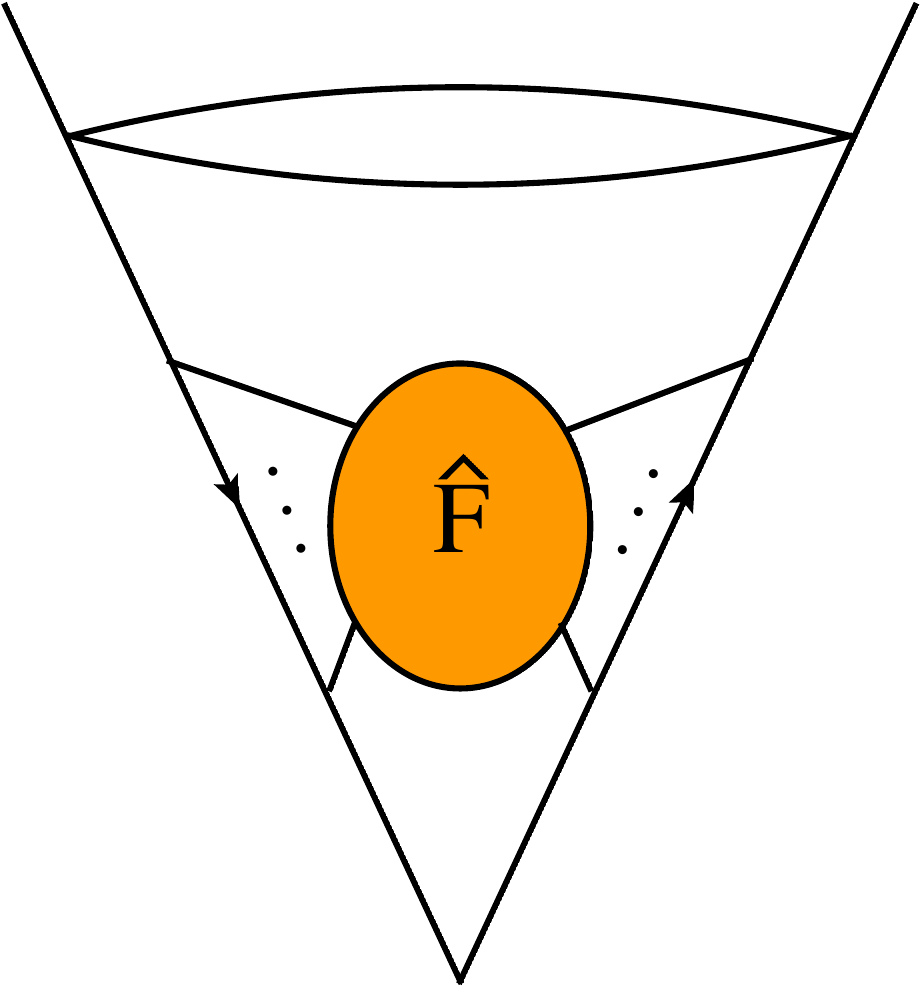}
\end{minipage}
\caption{Bethe-Salpeter equation at leading order}
\label{fig:BStrace}
\end{center}\end{figure}

Changing variables according to $s=L e^{\sigma '}$, $t=L e^{\tau '}$, where $L$ is an arbitrary length scale,
we get
\begin{equation}\label{traceBS1}
\begin{split}
&F(\sigma,\tau)=\sqrt{N}+\int_{-\infty}^{\sigma} d\sigma'\int_{-\infty}^{\tau} d\tau' 
\left(
 MN F(\sigma',\tau')P^{(B)}(\sigma',\tau')+\sqrt{MN} \hat{F}(\sigma',\tau')P^{(F)}(\sigma',\tau')\right),\\
&\hat{F}(\sigma,\tau)=\sqrt{M}+\int_{-\infty}^{\sigma} d\sigma'\int_{-\infty}^{\tau} d\tau' 
\left(
\sqrt{MN} F(\sigma',\tau') P^{(F)}(\sigma',\tau')+MN \hat{F}(\sigma',\tau') P^{(B)}(\sigma',\tau')
\right).
\end{split}
\end{equation}
with
\begin{equation}\begin{split}\label{PFPBsigma}
&P^{(F)}(\sigma,\tau)=-\left(\frac{2\pi}{k}\right)\frac{\Gamma(1/2-\epsilon)(\mu L)^{2\epsilon}}{2^{5/2-\epsilon}\pi^{3/2-\epsilon}}
\frac{\cos\theta/2}{\cos\varphi/2}\; 
(e^{\tau}\partial_{\sigma}+e^{\sigma}\partial_{\tau})\frac{e^{-\frac{1}{2}(\sigma+\tau)}e^{\epsilon(\sigma+\tau)}}{(\cosh(\sigma-\tau)+\cos\varphi)^{\frac{1}{2}-\epsilon}}\,,\\
&P^{(B)}(\sigma,\tau)=\left(\frac{2\pi}{k}\right)^2\frac{\Gamma^2(1/2-\epsilon)(\mu L)^{4\epsilon}}{2^{3-2\epsilon}\pi^{3-2\epsilon}}
\frac{\cos^2\theta/2}{\cos^2\varphi/2}\;\frac{\cos^2\varphi/2 
\;e^{2\epsilon(\sigma+\tau)}}{(\cosh(\sigma-\tau)+\cos\varphi)^{1-2\epsilon}}.
\end{split}\end{equation}
$F$ and $\hat F$ obey the differential equations
\begin{equation}\label{traceBS2}
\begin{split}
&\partial_{\sigma}\partial_{\tau}F(\sigma,\tau)=
MN F(\sigma,\tau)P^{(B)}(\sigma,\tau)+
\sqrt{MN} \hat{F}(\sigma,\tau)P^{(F)}(\sigma,\tau)\,,\\
&\partial_{\sigma}\partial_{\tau}\hat{F}(\sigma,\tau)=
\sqrt{MN} F(\sigma,\tau) P^{(F)}(\sigma,\tau)+
MN \hat{F}(\sigma,\tau) P^{(B)}(\sigma,\tau)\,,
\end{split}
\end{equation}
with boundary conditions $F(-\infty,\tau)\!=\!F(\sigma,-\infty)\!=\!\sqrt N$ and
$\hat F(-\infty,\tau)\!=\!\hat F(\sigma,-\infty)\!=\!\!~\sqrt M$.
Then we write $x=\sigma-\tau$ and $y=(\sigma+\tau)/2$ and obtain
\begin{equation}\label{traceBS3}
\begin{split}
&\left(\frac{1}{4}\partial^2_{y}-\partial^2_{x}\right)F(x,y)=
MN F(x,y)\tilde{P}^{(B)}(x,y)+
\sqrt{MN} \hat{F}(x,y)\tilde{P}^{(F)}(x,y)\,,\\
&\left(\frac{1}{4}\partial^2_{y}-\partial^2_{x}\right)
\hat{F}(x,y)=
\sqrt{MN} F(x,y) \tilde{P}^{(F)}(x,y)+
MN \hat{F}(x,y) \tilde{P}^{(B)}(x,y)\,,
\end{split}
\end{equation}
with
\small\begin{equation}\label{kery} 
\begin{split}
&\tilde{P}^{(F)}(x,y)=\left(\frac{2\pi}{k}\right)\frac{\Gamma\left(\frac 12-\epsilon\right)(\mu L)^{2\epsilon}}{(2\pi)^{3/2-\epsilon}}
\frac{\cos\frac \theta 2}{\cos\frac \varphi 2}e^{2\epsilon y}\!\! 
\left\{\frac{d}{dx}\!\left[\frac{\sinh x/2}{(\cosh x+\cos\varphi)^{\frac{1}{2}-\epsilon}}\right]-
\frac{\epsilon\,\cosh x/2}{(\cosh x+\cos\varphi)^{\frac{1}{2}-\epsilon}}\right\}\,,\\
&\tilde{P}^{(B)}(x,y)=\left(\frac{2\pi}{k}\right)^2\frac{\Gamma^2\left(\frac12-\epsilon\right)(\mu L)^{4\epsilon}}{(2\pi)^{3-2\epsilon}}
\!\!\frac{\cos^2\frac \theta 2}{\cos^2\frac\varphi 2}e^{4\epsilon y}\; 
\left\{\frac{(\cosh x+\cos\varphi)^{2\epsilon}}{2}-\frac{\sinh^2 x/2}{(\cosh 
x+\cos\varphi)^{1-2\epsilon}}\right\}.
\end{split}\end{equation}\normalsize

\subsection{General solution in $d=3$}\label{sec:sold3}
For $\epsilon=0$  equations \eqref{traceBS3}  can be decoupled  easily since
the kernels \eqref{kery}  are independent of $y$. Indeed,  by introducing 
\begin{equation} \label{HK}
\mathcal{H}(x,y)=F(x,y)+\hat{F}(x,y),\qquad\mathcal{K}(x,y)=F(x,y)-\hat{F}(x,y),
\end{equation}
\eqref{traceBS3} are equivalent to
\begin{equation}\label{traceBSHK}
\begin{split}
&\left(\frac{1}{4}\partial^2_{y}-\partial^2_{x}\right)\mathcal{H}(x,y)=
 \left(aW'(x)-a^2 W^2(x)+\frac{a^2}{2}\right)\mathcal{H}(x,y)\,,\\
&\left(\frac{1}{4}\partial^2_{y}-\partial^2_{x}\right)\mathcal{K}(x,y)=
\left(-aW'(x)-a^2 W^2(x)+\frac{a^2}{2}\right)\mathcal{K}(x,y).
\end{split}
\end{equation}
with
\begin{equation}\label{aaaaaaaaa}
a=\left(\frac{2\pi}{k}\right)\frac{\sqrt{MN}}{2^{3/2}\pi}
\frac{\cos\theta/2}{\cos\varphi/2}\,,
\end{equation}
and
\begin{equation}
W(x)=\frac{\sinh x/2}{(\cosh x+\cos\varphi)^{\frac{1}{2}}}\,.
\end{equation}
These equations can be solved using the separation variable method. Setting
\begin{equation}\label{sep}
\mathcal{H}(x,y)=h(y)\psi_{+}(x)\,,\qquad\mathcal{K}(x,y)=k(y)\psi_{-}(x)\,,
\end{equation}
we get 
\begin{equation}\begin{split}
&\partial^2_{y} h(y)=4(-E+\frac{a^2}{2})h(y)\,,\\
&\partial^2_{y} k(y)=4(-\tilde{E}+\frac{a^2}{2})k(y)\,,
\end{split}\end{equation}
and
\begin{equation}\label{traceBSx}
\begin{split}
&\left(-\partial^2_{x}+a^2W^2(x)-aW'(x)\right)\psi_{+}(x)=E\psi_{+}(x)\,,\\
&\left(-\partial^2_{x}+a^2W^2(x)+aW'(x)\right)\psi_{-}(x)=\tilde{E}\psi_{-}(x)\,.\end{split}
\end{equation}
The solution of the $y$ dependent equations is simply
\begin{equation}\begin{split}\label{soly}
& h(y)=C_1 e^{2\sqrt{-E+\frac{a^2}{2}}y}+C_2 e^{-2\sqrt{-E+\frac{a^2}{2}}y}\,,\\
&k(y)=C_3 e^{2\sqrt{-\tilde{E}+\frac{a^2}{2}}y}+C_4 
e^{-2\sqrt{-\tilde{E}+\frac{a^2}{2}}y}\,,
\end{split}\end{equation}
with $C_{1,2,3,4}$ constants which have to be fixed by imposing the boundary 
conditions as we will discuss in the following.
The $x-$dependent equations \eqref{traceBSx} can be seen as the two  Schroedinger
equations of a supersymmetric quantum mechanical system \cite{Cooper:1994eh}, therefore $E$
and $\tilde E$ are non-negative. In principle one could solve for these equations. 
However we are only interested to consider the case in which the edges of the 
cusp extend to infinity, i.e. $S$ and $T$ very large. In this limit $x\sim 0$ 
and $y$ is very large, thus we make  the ansatz $E=\tilde{E}=0$
\footnote{In the weak coupling limit for positive energy values the solutions of \eqref{soly} become oscillatory.}
and we set $\psi_+(0)=\psi_-(0)=1$ since they can be reabsorbed in the 
normalization constants $C_1,C_2,C_3,C_4$.

Using \eqref{sep}, \eqref{soly}  and \eqref{HK} we get
\begin{equation}\begin{split}\label{FhatF}
& F(0,y)=\frac{C_1+C_3}{2}e^{\sqrt2a y}+\frac{C_2+C_4}{2}e^{-\sqrt2a y}\,,\\
& \hat F(0,y)=\frac{C_1-C_3}{2}e^{\sqrt2ay}+\frac{C_2-C_4}{2}e^{-\sqrt2a y}.
\end{split}\end{equation}
We fix the constants $C_1,\dots,C_4$ by matching  the perturbative result.  In 
$d=3$  there are UV divergences coming from the integration regions close to the cusps. 
To isolate this divergence we  set $s_{min}=t_{min}=\delta$ which means 
$y_{min}=\ln\frac{\delta}{L}\equiv-\L$.
At tree level, obviously
\begin{equation}\label{bc}
F^{(0)}(0,-\L)=\sqrt{N},\qquad\text{and}\qquad\hat{F}^{(0)}(0,-\L)=\sqrt{M}.
\end{equation}
 Inserting these conditions in \eqref{FhatF} we obtain 
\begin{equation}\begin{split}\label{FhatF2}
& F(0,y)=\sqrt{N}e^{\sqrt2a(y+\L)}-\frac{C_2+C_4}{2}\sinh{\sqrt 2a(y+\L)}e^{\sqrt2a 
\L}\,,\\
& \hat{F}(0,y)=\sqrt{M}e^{\sqrt2a(y+\L)}-\frac{C_2-C_4}{2}\sinh{\sqrt2a(y+\L)}e^{\sqrt2a 
\L}\,.
\end{split}\end{equation}
The two remaining constants are determined  by matching the first order in the 
coupling (which is contained in $a$)  of our solution \eqref{FhatF2} with 
the first iteration of the Bethe-Salpeter equations \eqref{traceBS1} at the same 
order 
\begin{equation}\begin{split}
&F^{(1)}(\tau,\tau)=M\sqrt{N}\int_{-\L}^\tau d\sigma'\int_{-\L}^\tau d\tau' P^{F}(\sigma',\tau')\,,\\
&\hat{F}^{(1)}(\tau,\tau)=N\sqrt{M}\int_{-\L}^\tau d\sigma' \int_{-\L}^\tau d\tau' 
P^{F}(\sigma',\tau')\,,
\end{split}\end{equation}
which gives
\begin{equation}\begin{split}
&C_2=\frac{\sqrt{N}+\sqrt{M}}{2}(1-A)\,,\\
&C_4=\frac{\sqrt{N}-\sqrt{M}}{2}(1+A)\,,
\end{split}\end{equation}
where
\begin{equation}\label{Adef}
A=\lim_{\substack{y\to\infty\\\L\to\infty}}\frac{\sqrt{MN}}{\sqrt{2}a(y+\L)}\int_{-\L}^y d\sigma'
\int_{-\L}^y d\tau' 
P^{F}(\sigma',\tau')\,.
\end{equation}
In the appendix \ref{sec:appC} we compute this integral and we find  $A=1$, thus $C_2=0$ and $C_4=\sqrt N-\sqrt 
M$. Inserting this result in \eqref{FhatF2} we finally obtain
\begin{equation}\begin{split}\label{solphi}
\langle \mathcal{W}_+\rangle&=\frac{\sqrt{N}F+\sqrt{M}\hat{F}}{N+M}=\cosh 
\sqrt{2}a(y+\L)+\frac{2\sqrt{MN}}{N+M}\sinh\sqrt{2}a(y+\L)\,.\\
\langle \mathcal{W}_-\rangle&=\frac{\sqrt{N}F-\sqrt{M}\hat{F}}{N-M}=\cosh 
\sqrt{2}a(y+\L)\,.
\end{split}\end{equation}
In order to extract the cusp anomalous dimension at arbitrary $\varphi$ from the
result \eqref{solphi}, we have simply to recast them in a suitable form to single out the logarithmic divergence. 
Going back to the original (dimensionful) variables, we define
\begin{equation}\begin{split}
T=S=\Lambda_{IR}^{-1}=L e^{y}\quad &\rightarrow\quad y=-\log{L\Lambda_{IR}}=\log{\frac{T}{L}};\\
\delta=\Lambda_{UV}^{-1}=L e^{-\L}\quad &\rightarrow\quad 
\L=\log{L\Lambda_{UV}}=\log{\frac{L}{\delta}}\,,
\end{split}\end{equation}
where $\Lambda_{IR}$ is the natural IR cut-off (associated to the length of the cusp) and 
$\Lambda_{UV}=\frac{1}{\delta}$ is the UV cut-off (cutting-off the cusp, where ladders collapse). The length scale $L$, introduced previously for dimensional reason, will not play any role in the following. With the above definitions we get
\begin{equation}
(y+\L)=\log{\frac{T}{\delta}}=\log{\frac{\Lambda_{UV}}{\Lambda_{IR}}}\,.
\end{equation}

We can finally rewrite the expectation value of the cusped Wilson loop in the suggestive form as follows:
\begin{equation}\begin{split}\label{trdoppia2}
\langle \mathcal{W}_+\rangle=&\frac{(\sqrt{M}+\sqrt{N})^2}{2(M+N)}e^{\sqrt{2}a \log{\frac{\Lambda_{UV}}{\Lambda_{IR}}}}
+\frac{(\sqrt{M}-\sqrt{N})^2}{2(M+N)}e^{-\sqrt{2}a
\log{\frac{\Lambda_{UV}}{\Lambda_{IR}}}}\,,\\
\langle \mathcal{W}_-\rangle=&\frac12 e^{\sqrt{2}a \log{\frac{\Lambda_{UV}}{\Lambda_{IR}}}}
+\frac12e^{-\sqrt{2}a \log{\frac{\Lambda_{UV}}{\Lambda_{IR}}}}\,.
\end{split}\end{equation}
We have exactly reproduced the double-exponential structure found at two-loop in  
\cite{Griguolo:2012iq}: it comes from an all-order computation, in a particular limit, 
strongly supporting the mixing picture. 
In Section \ref{sec5} we will discuss how the cusp anomalous dimension is related to this result.

\subsection{The straight-line limit: the solution for $\epsilon \neq 0$}

We consider here the straight-line limit $\varphi=0$: remarkably the system
enjoys supersymmetry for any value of $\epsilon$ and the cusp anomalous 
dimension can be computed exactly.
To show this fact we first perform 
the change of variable $x\rightarrow i x'$ 
and $y\rightarrow \frac 12 y'$. Then using \eqref{HK}, eqs. \eqref{traceBS3} 
become
\begin{equation}
\begin{split}\label{HKeqphizerosusy}
&\Box\mathcal{H}(x',y')=
 \left[\vec{\nabla}W(x',y')\cdot\vec{\nabla}W(x',y')-\Box W(x',y')\right]\mathcal{H}(x',y')\,,\\
&\Box\mathcal{K}(x',y')=
\left[\vec{\nabla}W(x',y')\cdot\vec{\nabla}W(x',y')+\Box 
W(x',y')\right]\mathcal{K}(x',y')\,,
\end{split}
\end{equation}
with $\vec{\nabla}=(\partial_{x'},\partial_{y'})$, $\Box=\partial_{x'}^2+\partial_{y'}^2$ and
\begin{equation}\label{aepsilon}
W(x',y')=\frac{2^{\epsilon-1/2}a_{\epsilon}}{\epsilon} e^{\epsilon y'}
\cos^{2\epsilon}{\frac {x'}{2}}\,,
\;\;\;\;\;\;\;
a_{\epsilon}=\left(\frac{2\pi}{k}\right)\sqrt{MN}\frac{\Gamma(1/2-\epsilon)(\mu L)^{2\epsilon}}{(2\pi)^{3/2-\epsilon}}
\cos\theta/2\,.
\end{equation}
These equations are the Schroedinger  equations of the 
two bosonic sectors of a two-dimensional $\mathcal{N}=2$ supersymmetric 
quantum mechanics.
The wave function of the ground state can be exactly found and gives 
\begin{equation}\begin{split}\label{solution2}
\mathcal{H}(x,y)=& C_1 e^{-W(-ix',2y')}=C_1 e^{-\frac{2^{\epsilon-1/2}a_{\epsilon}}{\epsilon} e^{2\epsilon y}
\cosh^{2\epsilon}{\frac {x}{2}}}\,,\\
\mathcal{K}(x,y)=& C_2 e^{W(-ix',2y')}=C_2 e^{\frac{2^{\epsilon-1/2}a_{\epsilon}}{\epsilon} e^{2\epsilon y}
\cosh^{2\epsilon}{\frac {x}{2}}}\,,
\end{split}\end{equation}
with $C_1$ and $C_2$  normalization constants. Thus, using \eqref{HK},  one finds
\begin{equation}\begin{split}\label{solution2F}
F(x,y)=&\frac{C_1}{2} e^{-\frac{2^{\epsilon-1/2}a_{\epsilon}}{\epsilon} e^{2\epsilon y}
\cosh^{2\epsilon}{\frac {x}{2}}}+\frac{C_2}{2} e^{\frac{2^{\epsilon-1/2}a_{\epsilon}}{\epsilon} e^{2\epsilon y}
\cosh^{2\epsilon}{\frac {x}{2}}}\,,\\
\hat{F}(x,y)=&\frac{C_1}{2} e^{-\frac{2^{\epsilon-1/2}a_{\epsilon}}{\epsilon} e^{2\epsilon y}
\cosh^{2\epsilon}{\frac {x}{2}}}-\frac{C_2}{2} e^{\frac{2^{\epsilon-1/2}a_{\epsilon}}{\epsilon} e^{2\epsilon y}
\cosh^{2\epsilon}{\frac {x}{2}}}\,.
\end{split}\end{equation}
Here too we use the boundary conditions
\begin{equation}
F(0,-\infty)=\sqrt{N}\,,\quad \text{and} \qquad\hat{F}(0,-\infty)=\sqrt{M}\,,
\end{equation}
to fix the constants in \eqref{solution2F}, getting
\begin{equation}\begin{split}
C_1=&\sqrt{N}+\sqrt{M}\,,\\
C_2=&\sqrt{N}-\sqrt{M}\,.
\end{split}\end{equation}
The traced and supertraced operators for any $x$ and $y$ are
\begin{equation}\begin{split}\label{expphizero}
\langle \mathcal{W}^{\varphi=0}_+ \rangle=&\frac{\sqrt{N}F+\sqrt{M}\hat F}{N+M}
=\cosh V_{\epsilon}(x,y)+\frac{2\sqrt{MN}}{M+N}\sinh V_{\epsilon}(x,y)\,,\\
\langle \mathcal{W}^{\varphi=0}_- \rangle=&\frac{\sqrt{N}F-\sqrt{M}\hat F}{N-M}
=\cosh V_{\epsilon}(x,y)\,,
\end{split}\end{equation}
where
\begin{equation}
V_{\epsilon}(x,y)=-W(-ix,2y)=-\frac{2^{\epsilon-1/2}a_{\epsilon}}{\epsilon} e^{2\epsilon y}
\cosh^{2\epsilon}{\frac {x}{2}}\,.
\end{equation}
In order to consider the infinite cusped Wilson loop, 
recalling the original variables of the Bethe-Salpeter integrals,
we have to set $x,y=0$, then we have
\begin{equation}\begin{split}\label{phizero}
\langle \mathcal{W}_+^{\varphi=0}\rangle=&\frac{(\sqrt{M}+\sqrt{N})^2}{2(M+N)}e^{-\frac{2^{\epsilon-1/2}}{\epsilon} 
a_{\epsilon}}
+\frac{(\sqrt{M}-\sqrt{N})^2}{2(M+N)}e^{\frac{2^{\epsilon-1/2}}{\epsilon} 
a_{\epsilon}}\,,\\
\langle \mathcal{W}_-^{\varphi=0}\rangle=&\frac12 e^{-\frac{2^{\epsilon-1/2}}{\epsilon} 
a_{\epsilon}}
+\frac12e^{\frac{2^{\epsilon-1/2}}{\epsilon} 
a_{\epsilon}}\,.
\end{split}\end{equation}

\section{The determination of $\Gamma_{\text{cusp}}(\varphi)$}\label{sec5} 

Above we have seen that the quantum expectation value of our cusped Wilson loop
organizes itself as
a double exponential  in the limit \eqref{lim1} (see \eqref{trdoppia2} or
\eqref{phizero}).
Here  we want to extract the cusp anomalous dimension from the Bethe-Salpeter 
results. We find convenient to express the traced and supertraced operators  
$\mathcal{W}_{a}^{B}$, $a=\pm$, on a basis $\mathcal{W}_i^{B}$, $i=1,2$, 
whose elements  renormalize multiplicatively, i.e.
\begin{equation}\begin{split}
\mathcal{W}_i^{B}=&\,Z^{(i)}_{\text{cusp}}\;\mathcal{W}_i^{R}.
\end{split}\end{equation}
The two sets of  bare operators are related by a linear transformation depending
only on the ranks of 
the gauge groups
\begin{equation}\begin{split}
\mathcal{W}_{a}^{B}=&{\cal A}_{ai}\mathcal{W}_i^{B}.
\end{split}\end{equation}
From the explicit 
solution \eqref{trdoppia2} or \eqref{phizero} of the Bethe-Salpeter equation and
\eqref{mix1} 
one 
reads
\be\label{matrA}
\mathcal{A}=
\frac{1}{2(M+N)}
\left(
\begin{array}{cc}
 \left(\sqrt{M}+\sqrt{N}\right)^2 &
   \left(\sqrt{M}-\sqrt{N}\right)^2 \\
 M+N & M+N \\
\end{array}
\right)\,,
\ee
while the regulator dependent parts enter in the $Z^{(i)}_{\text{cusp}}$ (in this
discussion
we can neglect the additional divergences subtracted by  $Z_{open}$ since  they are
suppressed in the limit 
of large imaginary $\theta$).

In the $\varphi=0$ case one has 
\be
Z_{\text{cusp}}^{(1)}=e^{-\frac{2^{\epsilon-1/2}}{\epsilon}a_{\epsilon}}\,,
\;\;\;\;\;\;
Z_{\text{cusp}}^{(2)}=e^{\frac{2^{\epsilon-1/2}}{\epsilon}a_{\epsilon}}\,,
\ee
thus
\begin{equation}
\label{cuspida}
\begin{split}
\Gamma_{\text{cusp}}^{(1)}=&\mu\frac{\partial}{\partial\mu}\log(Z_{\text{cusp}}^{(1)})=-\frac{\sqrt{MN}}{\kappa}\cos\frac{\theta}{2}=-\sqrt{\hat{\lambda}_1\hat{\lambda}_2}\,,\\
\Gamma_{\text{cusp}}^{(2)}=&\mu\frac{\partial}{\partial\mu}\log(Z_{\text{cusp}}^{(2)})=\frac{\sqrt{MN}}{\kappa}\cos\frac{\theta}{2}=\sqrt{\hat{\lambda}_1\hat{\lambda}_2}\,,
\end{split}\end{equation}
where the definition \eqref{aepsilon}  for $a_{\epsilon}$ has been used and 
$ \hat \lambda_1$ and $\hat \lambda_2$  are the two effective coupling constants 
introduced in \eqref{lim1}. In the case of ABJM, where $N=M$, the two coupling
constants of 
course coincide: $\hat{\lambda}_1=\hat{\lambda}_2\equiv\hat{\lambda}$.

In order to extract the cusp anomalous dimension at arbitrary $\varphi$,
we have to consider the result \eqref{trdoppia2}. 
We first consider the ABJM case. As already announced in Section 
\ref{sec:sezionemaledetta} we observe a drastic simplification of the expectation value of the traced operator
with the disappearance of one of the two exponentials.  In other words  the trace renormalizes  multiplicatively.
Conversely the supertrace  stills  mixes with the trace.
Therefore, for $M=N$, we can associate  a cusp anomalous dimension, $\Gamma_{\text{cusp}}(\varphi,\hat{\lambda})$, directly to $\mathcal{W}_+$ though 
the usual definition \eqref{definitiongamma} and, using \eqref{aaaaaaaaa}, we find:
\begin{equation}
\label{face}
\langle \mathcal{W}_+\rangle^{ABJM}=e^{\sqrt{2}a
\log{\frac{\Lambda_{UV}}{\Lambda_{IR}}}}\qquad\Rightarrow\qquad 
\Gamma_{\text{cusp}}(\varphi)=-\sqrt{2}a=-\frac{N}{\kappa}
\frac{\cos\theta/2}{\cos\varphi/2}=-\frac{\hat{\lambda}}{\cos \frac{\varphi}{2}}.
\end{equation}

In the general situation, when $N\neq M$, the expectation value of both $\mathcal{W}_\pm$ contains a double exponential, and the coefficients of the divergent logarithms in the exponentials are identified with $\Gamma^{(1,2)}_{\text{cusp}}(\varphi)$:
\begin{equation} \label{cura}\begin{split}
\Gamma^{(1)}_{\text{cusp}}(\varphi)=&-\frac{\sqrt{\hat{\lambda}_1\hat{ \lambda}_2}}{\cos \frac{\varphi}{2}}\,,\  \ \ \ \ \ \ \ \
\Gamma^{(2)}_{\text{cusp}}(\varphi)=\frac{\sqrt{\hat{\lambda}_1\hat{ \lambda}_2}}{\cos \frac{\varphi}{2}},
\end{split}\end{equation}
as explained in more details in the case $\varphi=0$.  Moreover the result \eqref{cura} is  perfectly consistent with the 
limit  \eqref{cuspida}  and  for 
$\varphi=\theta=i \infty$, $\Gamma_{\text{cusp}}^{(1,2)}$ vanish as expected!

Some remarks on the results \eqref{trdoppia2} and \eqref{phizero} and the consequent form of  $\Gamma^{(i)}_{cusp}$ are now in order. First we analyze the structure of the exponentiation in the ABJ case: looking at our explicit calculation, one could expect that  the positive cusp anomalous dimension  dominates, while the negative  one gives a subleading contribution. On the other hand they appear on the same footing in our computations and, much more crucially, consistency with perturbative results requires the presence of  both of them.  However we believe that the  simple relation $\Gamma^{(1)}_{\text{cusp}}(\varphi)=-\Gamma^{(2)}_{\text{cusp}}(\varphi)$ implied by \eqref{cura} does not survive when the subleading corrections in 
$\theta$ are included.
  A second important observation concerns the actual functional form of 
  $\Gamma_{\text{cusp}}(\varphi)$. Let us concentrate for the moment on the $N=M$ case. 
  The final expression \eqref{face} is just the exponentiation of the one-loop result, namely 
  the leading cusp divergence undergoes to an abelian exponentiation in the ladder limit. This 
  result is completely different from the analogous 
  ${\cal N}=4$ SYM resummation, where an highly non-trivial function appears at this order, 
  even for $\varphi=~0$. The reason relies of course in the supersymmetric structure of the 
  effective Schroedinger equation but it has also a perturbative explanation: fermionic and 
  bosonic diagrams do not exponentiate in an abelian way by themselves and it is their 
  delicate balance that, order by order in the coupling constant, generates this nice behavior. 
  We have checked explicitly at three-loop in perturbation theory this fact. The $N\neq M$ 
  situation presents instead a slightly more involved structure: we have still an abelian-like 
  exponentiation at this order, but when expressed in terms of the two (scaled) 't Hooft 
  couplings $\hat{\lambda}_1,\,\,\hat{ \lambda}_2$ it appears through a square root of their 
  product. This is a further effect of the diagonalization process and at moment we do not 
  have a satisfying explanation from general principles. We stress that from the point of view 
  of the original CS level $k$ the exponentiation is one-loop as well. A third and, may be, 
  more interesting remark is related to the strong-coupling limit and the connection with string 
  theory. To be concrete, we shall consider the simpler case of ABJM: because of the 
  abelian-like exponentiation we do not have any non-trivial interpolation between weak 
  and strong-coupling and the scaling limit does not match the $\sqrt{\hat{\lambda}}$ 
  behavior of string theory. At variance with  ${\cal N}=4$ SYM the scaling limit does not seem 
  to commute with the strong-coupling limit, a fact that in four-dimensions was not expected 
  a priori (see the comments in the original computation \cite{Correa:2012nk}). We hope to 
  come back soon on this point when the subleading contributions will be computed.

\section{Conclusions}
\label{conclu}
In this paper we have studied a cusped Wilson loop in ${\cal N}=6$ Super Chern-Simons 
theory, constructed with lines that are 1/2 BPS. We have computed the associated cusp 
anomalous dimension in a scaling limit in which ladder diagrams dominate: because of the 
1/2 BPS character of the two halves, we have both bosonic and fermionic ladder exchanges 
and their resummation is encoded into a coupled Bethe-Salpeter equation. We have seen that 
it can be mapped into a supersymmetric Schroedinger equation whose ground state solution 
provide an exact expression for the cusp anomalous dimensions. Actually we found that, in the 
general $N\neq M$ case, the traced Wilson loop undergoes through a double-exponentiation, as first observed in \cite{Griguolo:2012iq}. 
This has been interpreted as an operator mixing under cusp renormalization: we have 
associated to the eigenvalues of the mixing matrix two independent cusp anomalous 
dimensions. The final result is very simple and the exponentiations are abelian, the cusp 
anomalous dimensions are one-loop exact up diagonalization. The strong-coupling limit  
is therefore trivial and we do not find consistency with string theory computation 
\cite{Forini:2012bb}: we argue that the scaling limit considered here does not commute with 
the strong-coupling limit. Concerning abelian exponentiation, a similar phenomenon has been 
observed recently \cite{Correa:2015wma} in studying ${\cal N}=4$ SYM cusped Wilson loops 
in $k$-symmetric representations: at large $N$ and $k$ planar diagrams dominate and the 
exponentiation is of abelian type.

The obvious follow-up of the present work is to take into account the subleading corrections 
to the scaling limit: in \cite{Henn:2012qz} a systematic approach to this computation has been 
developed in the ${\cal N}=4$ SYM case and it should be possible to perform an analogous 
investigation here. Preliminary results seem promising. It would be interesting to see if the 
supersymmetric structure we have found is preserved beyond leading order: in any case we 
expect a non-trivial modification of the relation 
$\Gamma^{(1)}_{\text{cusp}}(\varphi)=-\Gamma^{(2)}_{\text{cusp}}(\varphi)$. Another direction 
consists in checking the exponential structure at three-loop: the mixing we have observed here  
prescribes an exponentiation with definite group-dependent coefficients 
(see eq. (\ref{trdoppia2})), that appear to be the same both in the scaling limit and in the 
general two-loop result \cite{Griguolo:2012iq}. It would be of course nice to have a deeper 
understanding for the occurrence of the mixing coefficients: a closer look at the supersymmetric 
quantum mechanics discussed in \cite{Lee:2010hk}, where the 1/2 BPS line is obtained from a 
Higgsing procedure, should be probably useful for this task.

More ambitiously, one would like to approach the generalized cusp anomalous dimension in 
ABJ(M) theory from a general point of view, with the hope to obtain other all-order result by 
integrability or localization. In four-dimensions a particularly powerful approach has been pushed 
forward recently \cite{Gromov:2015dfa,Gromov:2016rrp}, applying to cusped Wilson loop the 
technique of the quantum spectral curve. Beautiful results have been obtained for the 
Bremsstrahlung function and the quark-anti-quark potential. It would be nice to extend this 
approach to ABJ(M) case, in which the quantum spectral curve has been already studied 
\cite{Cavaglia:2014exa}. It should be also possible to extend the TBA equations derived in 
\cite{Correa:2012hh,Drukker:2012de} in the three-dimensional context, taking advantage of the 
investigations presented in 
\cite{Bombardelli:2009xz,Gromov:2009at,Cavaglia':2013hva}.

\section*{Acknowledgements}

We thank Daniele Marmiroli for participating to the early stages of this work and 
Grisha Korchemsky for a very useful and enlightening discussion at GGI. These investigations have been 
supported in part by INFN and COST Action MP1210  "The String Theory Universe".

\vfill
\newpage

\appendix
	
\section{$\mathcal{N}=6$ three-dimensional Chern-Simons-matter theory}\label{sec:appA}

 \begin{figure}[h]
\centering
	\includegraphics[width=10cm]{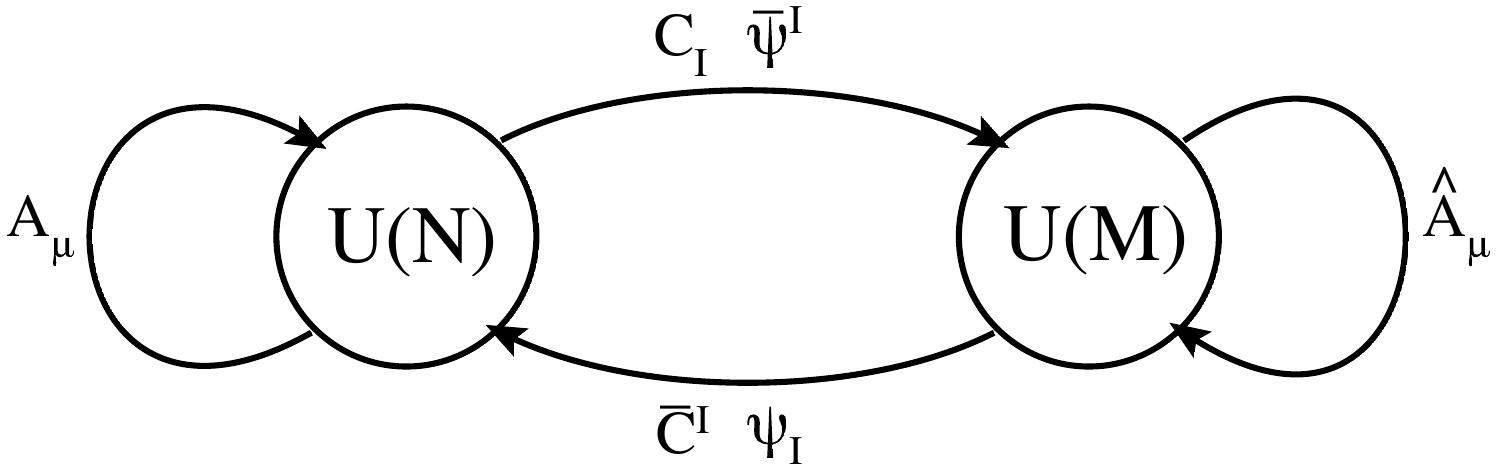}
	\caption{Quiver diagram for ABJ(M) theory.}
		\label{fig:quiver}
\end{figure}

The field content of the ABJ(M) theory can be schematically represented in the quiver 
in Figure \ref{fig:quiver}. 
The gauge sector consists of two gauge fields ${(A_\mu)_i}^j$ and 
${(\hat{A}_\mu)_{\hat{i}}}^{\hat{j}}$ belonging respectively to
the adjoint of $U(N)$ and $U(M)$. 
We denote by $i$, $\hat{i}$  the gauge indices in the
fundamental of the first and the second gauge group respectively.
The matter sector instead contains the complex fields
${(C_I)_i}^{\hat{j}}$ and ${(\bar{C}^I)_{\hat{i}}}^j$ as well as the fermions ${(\psi_{I})_{\hat{i}}}^j$ and ${(\bar{\psi}^I)_i}^{\hat{j}}$. 
The fields $(C, \bar\psi)$ transform in the $(\mathbf{N},\bar{\mathbf{M}})$ of the gauge group $U(N)\times U(M)$ 
while the pair $(\bar{C},\psi)$ lives in the $(\bar{\mathbf{N}},\mathbf{M})$.
The additional capitol index $I=1,2,3,4$ belongs to the R-symmetry group 
$SU(4)$. 
The ABJ(M) action is
\begin{equation}\label{abjmactiontot}
S_{\text{ABJ(M)}}=S_{\text{CS}}+S_{\text{gf}}+S_{\text{Matter}}+S^{F}_{\text{int}}+S^{B}_{\text{int}}\,,
\end{equation}
where
\begin{equation}\begin{split}
S_{\text{CS}}=&-i\frac{\kappa}{4\pi}\int d^3x \epsilon^{\mu\nu\rho}\left[\text{Tr}(A_\mu\partial_\nu A_\rho+\tfrac 23 i A_\mu A_\nu A_\rho)-
\text{Tr}(\hat{A}_\mu\partial_\nu \hat{A}_\rho+\tfrac 23 i \hat{A}_\mu \hat{A}_\nu 
\hat{A}_\rho)\right],\\
S_{\text{gf}}=&\frac{\kappa}{4\pi}\int d^3x 
\left[\tfrac 1\xi \text{Tr}\left(\partial_\mu A_\mu\right)^2+\text{Tr}\left(\partial_\mu \bar{c} D_\mu c\right)
-\tfrac 1\xi \text{Tr}(\partial_\mu \hat{A}_\mu)^2+\text{Tr}\left(\partial_\mu \bar{\hat{c}} D_\mu 
\hat{c}\right)\right],\\
S_{\text{Matter}}=&\int d^3x \left[\text{Tr}\left(D_\mu C_I D^\mu 
\bar{C}^I\right)+i \text{Tr}\left(\bar{\psi}^I\slashed{D}\psi_I\right)\right],
\end{split}\end{equation}
\normalsize
and
\begin{equation}\begin{split}
S^{F}_{\text{int}}=&-\frac{2\pi i}{\kappa}\int d^3x
\left[ \text{Tr}(\bar{C}^I C_I \psi_J \bar{\psi}^J)-\text{Tr}( C_I 
\bar{C}^I\bar{\psi}^J\psi_J)+2\text{Tr}( C_I \bar{C}^J\bar{\psi}^I\psi_J)\right.\\
&\qquad\qquad\left. -2\text{Tr}(\bar{C}^I C_J \psi_I 
\bar{\psi}^J)-\epsilon_{IJKL}\text{Tr}(\bar{C}^I\bar{\psi}^J\bar{C}^K\bar{\psi}^L)+\epsilon^{IJKL}\text{Tr}(C_I\psi_J C_K 
\psi_L)\right],\\
S^{B}_{\text{int}}=&-\frac{4\pi^2}{3\kappa^2}\int d^3x
\left[ \text{Tr}(C_I\bar{C}^I C_J\bar{C}^J C_K\bar{C}^K)+\text{Tr}(\bar{C}^I C_I \bar{C}^J C_J \bar{C}^K C_K)\right. \\
&\qquad\qquad\quad\qquad\left. +4\text{Tr}(C_I\bar{C}^J C_K\bar{C}^I C_J\bar{C}^K)-6 \text{Tr}(C_I\bar{C}^J C_J\bar{C}^I 
C_K\bar{C}^K)\right],
\end{split}\end{equation}
where $\epsilon^{1234}=\epsilon_{1234}=1$ and $\kappa$ is the Chern-Simons level.
The matter covariant derivatives are defined as
\begin{equation}\begin{split}
D_\mu C_I=&\partial_\mu C_I+i(A_\mu C_I-C_I\hat{A}_\mu),\qquad
D_\mu \bar{C}^I=\partial_\mu\bar{C}^I-i(\bar{C}^I A_\mu - 
\hat{A}_\mu\bar{C}^I),\\
D_\mu \psi_I=&\partial_\mu \psi_I+i(\hat{A}_\mu \psi_I-\psi_I A_\mu),\qquad\;
D_\mu \bar{\psi}^I=\partial_\mu\bar{\psi}^I-i(\bar{\psi}^I \hat{A}_\mu-A_\mu \bar{\psi}^I).
\end{split}\end{equation}

\paragraph{Propagators and bilinears} $\\$

The position-space propagators are obtained  from those in momentum space (see {\it e.g.} \cite{Drukker:2008zx}) 
by means of  the following master integral
\be
\int \frac{d^{3-2\epsilon} p}{(2\pi)^{3-2\epsilon}} \frac{e^{i p\cdot x}}{(p^{2})^{s}}=\frac{\Gamma\left(\frac{3}{2}-s-\epsilon\right)}{4^{s} \pi^{\frac{3}{2}-\epsilon}\Gamma(s)}\frac{1}{(x^{2})^{\frac{3}{2}-s-\epsilon}}.
\ee
In the Landau gauge, we have the following propagators
\be
\begin{split}\label{propABJM}
\langle (A_{\mu})_{i}^{\ j}(x)  (A_{\nu})_{k}^{\  l} (y)\rangle=&\delta_{i}^{l}\delta_{k}^{j}\left(\frac{2\pi i}{\kappa}\right)\epsilon_{\mu\nu\rho}\partial^{\rho}_{x}D(x-y),\\
\langle (\hat A_{\mu})_{\hat i}^{\ \hat j}(x)  (\hat A_{\nu})_{\hat k}^{\ \hat l} (y)\rangle=&-\delta_{\hat i}^{\hat l}\delta^{\hat j}_{\hat k}\left(\frac{2\pi i}{\kappa}\right)\epsilon_{\mu\nu\rho}\partial^{\rho}_{x}D(x-y).\\
\langle (C_{I})_{i}^{\ \hat j}(x)  (\bar C^{J})_{\hat k}^{\  l} (y)\rangle=& \delta^{J}_{I}~\delta_{i}^{l}~\delta_{\hat k}^{\hat j} D(x-y),\\
\langle(\psi_{I})_{\hat i}^{\  j}(x)(\bar\psi^{J})_{ k}^{\ \hat l}(y)\rangle=&\delta^{J}_{I}~\delta_{\hat i}^{\hat l}~\delta_{ k}^{ j} i\gamma^{\mu}\partial_{\mu}D(x-y)\,,
\end{split}
\ee
where
\begin{equation}
D(x-y)\equiv \frac{\Gamma\left(\frac{1}{2}-\epsilon\right)}{4 
\pi^{\frac{3}{2}-\epsilon}}\frac{1}{((x-y)^{2})^{\frac{1}{2}-\epsilon}}\,.
\end{equation}
Computing the fermionic diagram contributing  to the Wilson loop 
we have to deal with the bilinear $\eta \gamma \bar \eta$.
Its expression in terms of the position along the line is \cite{Griguolo:2012iq}
\be
\label{rg6}
\begin{split}
(\eta_{2}\gamma^{\mu}\bar\eta_{1})=&-\frac{2}{(\eta_{1}\bar\eta_{2})}\left[\frac{\dot{x_{1}}^{\mu}}{|\dot x_{1}|}+\frac{\dot{x_{2}}^{\mu}}{|\dot{x}_{2}|}-i\frac{\dot{x_{2}}^{\lambda}}{|\dot{x}_{2}|}
\frac{\dot{x_{1}}^{\nu}}{|\dot x_{1}|}\epsilon_{\lambda\nu}^{\ \  \ \mu}
\right]\,,
\end{split}
\ee
where 1 and 2 denote two different points of the contour.

For our specific circuit all the possible products between $\eta$'s are the 
following
\begin{align}
&\eta_{1}\bar\eta_{1}=\eta_{2}\bar\eta_{2}=2 i,\ \ \ \ \eta_{1}\bar\eta_{2}=\eta_{2}\bar\eta_{1}=2 i\cos\frac{\varphi}{2},\nonumber\\
&\eta_{2}\eta_{1}=-2 i \sin\frac{\varphi}{2},\ \ \ \  \bar\eta_{1}\bar\eta_{2}=2 i \sin\frac{\varphi}{2}.
\end{align}
Here the indices $1$ and $2$ label the two different edges of the cusp.

\section{Useful integral}\label{sec:appC}

 In the following we want to compute the quantity $A$ appearing in the Section \ref{sec:sold3} and defined by
 \begin{equation}\label{Adef2}
A=\lim_{\substack{y\to\infty\\\L\to\infty}}\frac{\sqrt{MN}}{\sqrt{2}a(y+\L)}
\int_{-\L}^y d\tau' \int_{-\L}^y d\sigma'
P^{(F)}(\sigma',\tau').
\end{equation}
It is convenient to have different upper bounds in the integrals. Using the definition of $P^{(F)}$ given by \eqref{PFPBsigma}, we have:
\small
\begin{equation}\begin{split}
\int_{-\L}^\tau d\tau' \int_{-\L}^\sigma d\sigma'
P^{(F)}(\sigma',\tau')&
=-\frac{\cos\theta/2}{\sqrt{2}\kappa}\frac{1}{\cos\varphi/2}
\int_{-\L}^{\tau}d\tau'\int_{-\L}^{\sigma}d\sigma'\frac{d}{d \sigma'}
\left(\frac{e^{\frac{1}{2}(\tau'-\sigma')}}{(\cosh(\sigma'-\tau')+\cos\varphi)^{\frac{1}{2}}}\right)\\
&=-\frac{\cos\theta/2}{\sqrt{2}\kappa}\frac{1}{\cos\varphi/2}\biggl(I(\sigma,\tau)-I(-\L,\tau)\biggl)\,,
\end{split}\end{equation}
\normalsize
where we have computed the first integral using the total derivative and where we have 
defined:
\begin{equation}
I(\sigma,\tau)\equiv \int_{-\L}^{\tau}d\tau'
\frac{e^{\frac{1}{2}(\tau'-\sigma)}}{(\cosh(\sigma-\tau')+\cos\varphi)^{\frac{1}{2}}}.
\end{equation}
We perform the change of variable $z=e^{\sigma-\tau'}$ and solve the first 
integral:
\begin{equation}\begin{split}
I(\sigma,\tau)=&\sqrt{2}\int_{-\L}^{\tau}d\tau'\frac{1}{\left[e^{2(\sigma-\tau')}+2e^{(\sigma-\tau')}\cos\varphi+1\right]^{1/2}}\\
=&-\sqrt{2}\int_{e^{(\sigma+\L)}}^{e^{(\sigma-\tau)}}\frac{dz}{z}\frac{1}{\left[z^2+2z\cos\varphi+1\right]^{1/2}}\\
=&-\sqrt{2}\biggr[-\log\left(\frac{1+z\cos\varphi+\sqrt{z^2+2z\cos\varphi+1}}{z}\right)\biggr]_{e^{(\sigma+ \L)}}^{e^{(\sigma-\tau)}}.
\end{split}\end{equation}
The second contribution is:
\begin{equation}
I(-\L,\tau)=-\sqrt{2}\biggr[-\log\left(\frac{1+z\cos\varphi+\sqrt{z^2+2z\cos\varphi+1}}{z}\right)\biggr]_{1}^{e^{-(\tau+\L)}}\,.
\end{equation}
Summing up, we obtain:
\begin{equation}
I(\sigma,\tau)-I(-\L,\tau)=\sqrt{2}\biggr[G(\sigma-\tau)-G(\sigma+\L)-G(-\tau-\L)+G(0)\biggr]\,,
\end{equation}
where
\begin{equation}
G(x)=\log\left(1+e^{x}\cos\varphi+\sqrt{e^{2x}+2e^{x}\cos\varphi+1}\right)\,.
\end{equation}
Now setting $\tau=\sigma=y$, we have
\begin{equation}
I(y,y)-I(-\L,y)=\sqrt{2}\biggr[2G(0)-G(-y-\L)-G(y+\L)\biggr]\,.  
\end{equation}
For large $y$ and $\L$, we can write the following 
expansion:
\begin{equation}
I(y,y)-I(-\L,y)\simeq 
-\sqrt{2}(y+\L)+\text{const}+\mathcal{O}(e^{-(y+\L)})\,.
\end{equation}
Therefore
\begin{equation}
\sqrt{MN}\int_{-\L}^y d\tau' \int_{-\L}^y d\sigma'
P^{(F)}(\sigma',\tau')\simeq 
\frac{\cos\theta/2}{\cos\varphi/2}\frac{\sqrt{MN}}{\kappa}(y+\L)=\sqrt{2}a(y+L)\,.
\end{equation}
Recalling the definition \eqref{Adef2}, we obtain:
\begin{equation}
A=1\,.
\end{equation}

\end{document}